\documentclass[%
 reprint,
 amsmath,amssymb,
 aps,
 pre,
 longbibliography
]{revtex4-2}

\usepackage[pdftex]{graphicx} % for arXiv
\usepackage{dcolumn}% Align table columns on decimal point
\usepackage{bm}% bold math
\usepackage[T1]{fontenc}
\usepackage[pdftex]{color, hyperref}% for arXiv add hypertext capabilities
\hypersetup{
colorlinks = true,
linkcolor = blue,
citecolor = blue
}
\usepackage[normalem]{ulem}

\usepackage{color}

\begin{document}

% \preprint{APS/123-QED}

\title{Universal bounds on the performance of information-thermodynamic engine}% Force line breaks with \\
% \thanks{A footnote to the article title}%

\author{Tomohiro Tanogami$^1$}
\author{Tan Van Vu$^{2}$}
\author{Keiji Saito$^{3}$}
\affiliation{
$^1$ Department of Earth and Space Science, Osaka University, Osaka 560-0043, Japan\\ 
$^2$ Analytical quantum complexity RIKEN Hakubi Research Team, RIKEN Center for Quantum Computing (RQC), 2-1 Hirosawa, Wako, Saitama 351-0198, Japan\\
$^3$ Department of Physics, Kyoto University, Kyoto 606-8502, Japan}%Lines break automatically or can be forced with \\

% \collaboration{MUSO Collaboration}%\noaffiliation

% \author{Charlie Author}
 % \homepage{http://www.Second.institution.edu/~Charlie.Author}
% \affiliation{
 % Second institution and/or address\\
 % This line break forced% with \\
% }%
% \affiliation{
 % Third institution, the second for Charlie Author
% }%
% \author{Delta Author}
% \affiliation{%
%  Authors' institution and/or address\\
%  This line break forced with \textbackslash\textbackslash
% }%

% \collaboration{CLEO Collaboration}%\noaffiliation

\date{\today}% It is always \today, today,
             %  but any date may be explicitly specified
\begin{abstract}
% \begin{description}
% \item[Usage]
% Secondary publications and information retrieval purposes.
% \item[PACS numbers]
% May be entered using the \verb+\pacs{#1}+ command.
% \item[Structure]
% You may use the \texttt{description} environment to structure your abstract;
% use the optional argument of the \verb+\item+ command to give the category of each item.
% \end{description}
%%%%%% Introduction
%%%%%% Details of motivation/previous studies
%%%%%% Research question
%%%%%% Result
We investigate fundamental limits on the performance of information processing systems from the perspective of information thermodynamics.
%%%%%% Details of the result
We first extend the thermodynamic uncertainty relation (TUR) to a subsystem.
Specifically, for a bipartite composite system consisting of a system of interest $X$ and an auxiliary system $Y$, we show that the relative fluctuation of an arbitrary current for $X$ is lower bounded not only by the entropy production associated with $X$ but also by the information flow between $X$ and $Y$.
As a direct consequence of this \textit{bipartite} TUR, we prove universal trade-off relations between the output power and efficiency of an information-thermodynamic engine in the fast relaxation limit of the auxiliary system.
In this limit, we further show that the Gallavotti-Cohen symmetry is satisfied even in the presence of information flow.
This symmetry leads to universal relations between the fluctuations of information flow and entropy production in the linear response regime.
We illustrate these results with simple examples: coupled quantum dots and coupled linear overdamped Langevin equations.
Interestingly, in the latter case, the equality of the bipartite TUR is achieved even far from equilibrium, which is a very different property from the standard TUR.
%%%%%% Significance/impact
Our results will be applicable to a wide range of systems, including biological systems, and thus provide insight into the design principles of biological systems.
%%%%%% Implications
\end{abstract}

\pacs{Valid PACS appear here}% PACS, the Physics and Astronomy
                             % Classification Scheme.
%\keywords{Suggested keywords}%Use showkeys class option if keyword
                              %display desired

\maketitle
\section{Introduction}
%%%%%% Significance of information processing in biological systems
Biological systems maintain their functions by acquiring or using information about fluctuating environments.
For example, \textit{E.~coli} regulates its flagellar motors by processing information about external ligand concentrations to adapt to the environment~\cite{barkai1997robustness,alon1999robustness,tu2008modeling,lan2012energy,ito2015maxwell,mattingly2021escherichia}.
A gene network senses a sudden increase in protein concentration and then suppresses mRNA transcription to maintain protein levels~\cite{shen2002network,lee2002transcriptional,karlebach2008modelling,otsubo2018information}.
While these systems rely on a negative feedback mechanism that suppresses intrinsic noise by using information about fluctuating environments, some molecular machines can even convert information into output work.
Such examples include $\mathrm{F}_\mathrm{o}\mathrm{F}_1$-ATP synthase, where $\mathrm{F}_1$ motor converts energy and information provided by $\mathrm{F}_\mathrm{o}$ motor into the synthesis of ATP molecules~\cite{lathouwers2022internal,leighton2023inferring,grelier2023unlocking}.
To elucidate the general design principles underlying biological systems, it is necessary to investigate the fundamental limits on the performance of such information processing systems.

%%%%%% Stochastic thermodynamics 
Stochastic thermodynamics has revealed various fundamental limits to the thermodynamic aspects of such fluctuating mesoscale systems~\cite{sekimoto2010stochastic,seifert2012stochastic,van2015ensemble,peliti2021stochastic}.
For example, the thermodynamic uncertainty relation (TUR) states that suppressing the relative fluctuations of an arbitrary time-integrated current $\hat{\mathcal{J}}_\mathcal{T}$ necessarily involves a thermodynamic cost~\cite{barato2015thermodynamic,horowitz2020thermodynamic,shiraishi2021optimal}:
\begin{align}
\dfrac{\mathrm{Var}[\hat{\mathcal{J}}_\mathcal{T}]}{\langle \hat{\mathcal{J}}_\mathcal{T}\rangle^2}\ge\dfrac{2}{\Delta \sigma},
\label{standard TUR}
\end{align}
where $\langle \hat{\mathcal{J}}_\mathcal{T}\rangle$ and $\mathrm{Var}[\hat{\mathcal{J}}_\mathcal{T}]$ denote the mean and variance of $\hat{\mathcal{J}}_\mathcal{T}$, and $\Delta \sigma$ denotes the total entropy production up to time $\mathcal{T}$.
While the validity of TUR in its original form (\ref{standard TUR}) is limited to steady-state currents in Markov jump processes and overdamped Langevin dynamics, TUR-type inequalities even revealed that there is a fundamental limit to the performance of a thermodynamic heat engine.
Specifically, a heat engine with a finite output power cannot achieve the Carnot efficiency as long as the fluctuation of the output power is finite~\cite{shiraishi2016universal,pietzonka2018universal,holubec2018cycling,holubec2021fluctuations}.
Furthermore, for a stationary cross-transport system with input and output currents, which can be regarded as fuel (positive entropy) and load (negative entropy), respectively, the input-output fluctuation inequalities hold in the linear response regime~\cite{saryal2021universal,shiraishi2022there}.
These inequalities state that the fluctuation of the output current is smaller than that of the input current, while the relative fluctuation of the output current is larger than that of the input current.

%%%%%% Information thermodynamics & Research question
In this paper, we aim to find similar fundamental limits for information processing systems, in particular for an information-thermodynamic engine that converts information into output work.
Information thermodynamics, which is essentially stochastic thermodynamics for subsystems, is a thermodynamic framework for information flow between two interacting subsystems, either autonomous or nonautonomous~\cite{parrondo2015thermodynamics}.
This theory reveals that the information flow between subsystems can significantly affect the thermodynamic constraints of each subsystem.
While information thermodynamics has its origins in the thought experiment of Maxwell's demon, it has recently been applied to information processing at the cellular level in biological systems~\cite{barato2014efficiency,sartori2014thermodynamic,ito2015maxwell,hartich2016sensory,amano2022insights} and even to fully developed fluid turbulence~\cite{tanogami2022information}.

%%%%%% Results
Here, we consider a composite system consisting of a system of interest $X$ and an auxiliary system $Y$, described by continuous-time Markov jump processes or diffusion processes with only even variables and parameters under time reversal.
Our main results can be summarized as follows.
\begin{itemize}
	\item[(i)] \textit{Bipartite TUR}.---We first extend the standard TUR (\ref{standard TUR}) to a subsystem.
	For arbitrary time-integrated current $\hat{\mathcal{J}}_\mathcal{T}$ for $X$ with arbitrary observation time $\mathcal{T}$, we prove that [cf.~Ineq.~\eqref{bipartite TUR}]
	\begin{align}
	\dfrac{\mathrm{Var}[\hat{\mathcal{J}}_\mathcal{T}]}{\langle \hat{\mathcal{J}}_\mathcal{T}\rangle^2}\ge\dfrac{2(1+\delta_{\mathcal{J}})^2}{\Delta S^X_{\mathrm{tot}}-\Delta I^X}.
	\end{align}
	Here, $\Delta S^X_{\mathrm{tot}}$ denotes the entropy production associated with $X$, and $\Delta I^X$ denotes the time-integrated information flow, which is the amount of information exchanged with the auxiliary system $Y$.
	The additional term $\delta_{\mathcal{J}}$ reflects the contribution of the interaction with $Y$.
	This \textit{bipartite} TUR states that the relative fluctuation of the current for the subsystem $X$ is lower bounded not only by the entropy production associated with $X$, but also by the information transfer between $X$ and $Y$.
	In particular, if $Y$ evolves much faster than $X$, we can further show that $\delta_{\mathcal{J}}\rightarrow0$ in the steady state.
	In this case, the bipartite TUR gives a tighter bound than the standard TUR (\ref{standard TUR}).
	While here we derive the bipartite TUR in the steady state, this relation is valid even for systems under arbitrary time-dependent driving from arbitrary initial states (see Appendix~\ref{Bipartite TUR for overdamped Langevin equations}).
	\item[(ii)] \textit{Trade-off relations}.---As a consequence of the bipartite TUR, we show that there are fundamental limits on the performance of an information-thermodynamic engine.
	When the system of interest $X$ acts as a steady-state information-thermodynamic engine, its performance can be quantified, e.g., by the negative entropy production rate $|\dot{S}^X_{\mathrm{env}}|$ in the environment and the information-thermodynamic efficiency $\eta^X_S:=|\dot{S}^X_{\mathrm{env}}|/|\dot{I}^X|$, which quantifies how efficiently the engine converts information into the negative entropy production.
	In the typical case where the auxiliary system $Y$ evolves much faster than the engine $X$, we prove universal trade-off relations between $|\dot{S}^X_{\mathrm{env}}|$ and $\eta^X_S$ [cf.~Ineqs.~\eqref{trade-off between entropy production and efficiency} and \eqref{trade-off between entropy production and efficiency via D_I}]:
	\begin{align}
	|\dot{S}^X_{\mathrm{env}}|\le D_S\dfrac{1-\eta^X_S}{\eta^X_S},
	\end{align}
	and
	\begin{align}
	|\dot{S}^X_{\mathrm{env}}|\le D_I\eta^X_S(1-\eta^X_S),
	\end{align}
	where $D_S$ and $D_I$ denote the fluctuation of the stochastic medium entropy production and the time-integrated stochastic information flow, respectively.
	These inequalities state that an information engine with a finite negative entropy production rate cannot achieve $\eta^X_S=1$ as long as the fluctuations $D_S$ and $D_I$ are finite.
	In order to achieve a finite negative entropy production rate with $\eta^X_S=1$, the fluctuations $D_S$ and $D_I$ must diverge.
	\item[(iii)] \textit{Gallavotti-Cohen symmetry}.---In addition to the TURs, the Gallavotti-Cohen symmetry~\cite{gallavotti1995dynamical,kurchan1998fluctuation,lebowitz1999gallavotti} also provides important information about fluctuations of currents.
	For the scaled cumulant generating function $\mu(\lambda_S,\lambda_I)$ with the counting fields $\lambda_S$ and $\lambda_I$ for the stochastic medium entropy production and the time-integrated stochastic information flow, we prove that the following Gallavotti-Cohen symmetry holds in the fast relaxation limit of $Y$ [cf.~Eq.~\eqref{GC symmetry}]:
	\begin{align}
	\mu(\lambda_S,\lambda_I)=\mu(-\lambda_S-1,-\lambda_I-1).
	% \label{GC symmetry}
	\end{align}
	\item[(iv)] \textit{Input-output fluctuation inequalities}.---As a direct consequence of the Gallavotti-Cohen symmetry, we show that the input-output fluctuation inequalities hold even in the case where information flow is regarded as an input or output current.
	That is, in the linear response regime where $X$ acts as a steady-state information-thermodynamic engine, we prove that [cf.~Ineqs.~\eqref{Input-output inequality_1} and \eqref{Input-output inequality_2}]
	\begin{align}
	D_S&\le D_I,\\
	\dfrac{D_I}{(\dot{I}^X)^2}&\le \dfrac{D_S}{(\dot{S}^X_{\mathrm{env}})^2}.
	\end{align}
	These inequalities state that the fluctuation of the output current (negative entropy production) is smaller than that of the input current (information flow), while the relative fluctuation of the output current is larger than that of the input current.	
\end{itemize}
We illustrate these results with two simple examples: coupled quantum dots and coupled linear overdamped Langevin equations.
Interestingly, the latter provides an example where the equality of the bipartite TUR is achieved even far from equilibrium.
This is in contrast to the standard TUR (\ref{standard TUR}), where the equality is guaranteed only in the near-equilibrium limit~\cite{hasegawa2019uncertainty,shiraishi2021optimal}.
While the bipartite TUR is generally not valid for systems with broken time-reversal symmetry, such as underdamped Langevin dynamics~\cite{van2019uncertainty,lee2019thermodynamic,fischer2020free,lee2021universal,fu2022thermodynamic,dechant2022bounds,pietzonka2022classical}, many relevant biological systems are often described by continuous-time Markov jump processes or diffusion processes with only even variables and parameters under time reversal.
Therefore, these results will be applicable to a wide range of systems, including biological systems, and thus shed new light on our understanding of the design principles of biological systems.

%%%%%% Organization of the paper
This paper is organized as follows.
In Sec.~\ref{Setup}, we introduce important information-theoretic quantities and briefly review the framework of information thermodynamics in a general setup.
In Sec.~\ref{Bipartite TUR}, we describe the bipartite TUR, which is the first main result of this paper.
The detailed derivation of the bipartite TUR is presented in Sec.~\ref{Derivation of the bipartite TUR}.
In Sec.~\ref{Fast relaxation limit for Y}, we show that the bipartite TUR reduces to the form of the standard TUR if the auxiliary system evolves much faster than the system of interest.
We discuss the equality condition of the bipartite TUR in Sec.~\ref{Equality condition}.
In Sec.~\ref{Trade-off relations}, we show that the bipartite TUR gives universal bounds on the performance of an information-thermodynamic engine, which is the second main result of this paper.
In Sec.~\ref{Gallavotti-Cohen symmetry}, as the third main result of this paper, we prove that the Gallavotti-Cohen symmetry holds even in the presence of information flow in the fast relaxation limit of the auxiliary system.
As a corollary to this symmetry, we show that the input-output fluctuation inequalities are valid even in the case where information flow is regarded as an input or output current in Sec.~\ref{Input-output fluctuation inequalities}.
In Sec.~\ref{Examples}, we illustrate our results with two examples.
In Sec.~\ref{Concluding remarks}, we conclude this paper with some remarks.

\section{Setup\label{Setup}}
%&&&&& Bipartite Markov jump
We consider a composite system that consists of two subsystems, $X$ (system of interest) and $Y$ (auxiliary system), whose time evolution is described by Markov jump processes or overdamped Langevin equations.
Let $x_t$ and $y_t$ be the states of $X$ and $Y$ at time $t$, respectively.
We assume that the system satisfies the bipartite property: the transition probability $p(x_{t+dt},y_{t+dt}|x_t,y_t)$ satisfies
\begin{align}
p(x_{t+dt},y_{t+dt}|x_t,y_t)=p(x_{t+dt}|x_t,y_t)p(y_{t+dt}|x_t,y_t)
\end{align}
for $dt\rightarrow0^+$.
This property means that $X$ and $Y$ do not jump simultaneously in the case of Markov jump processes and that the noises acting on $X$ and $Y$ are uncorrelated in the case of diffusion processes.
In this paper, we focus mainly on Markov jump processes, while the extension to the overdamped Langevin case is straightforward.
Let $p_t(x,y)$ be the probability of state $(x,y)$ at time $t$.
The time evolution of $p_t(x,y)$ is described by the master equation:
\begin{align}
\partial_tp_t(x,y)=\sum_{x'}w^y_{xx'}p_t(x',y)+\sum_{y'}w^{yy'}_xp_t(x,y'),
\label{master equation}
\end{align}
where $w^y_{xx'}$ ($w^{yy'}_x$) is the time-independent transition rate from state $(x',y)$ to $(x,y)$ ($(x,y')$ to $(x,y)$) with $w^y_{xx}=-\sum_{x'(\neq x)}w^y_{x'x}$ ($w^{yy}_x=-\sum_{y'(\neq y)}w^{y'y}_x$).
The rate matrix is assumed to be irreducible to ensure the uniqueness of the stationary distribution $p_{\mathrm{ss}}(x,y)$.
Note that $X$ and $Y$ can affect each other's transition rates, although they cannot jump simultaneously.

\subsection{Information-theoretic quantities}
We introduce important information-theoretic quantities.
The strength of the correlation between $X$ and $Y$ can be quantified by the mutual information~\cite{cover1999elements}:
\begin{align}
I[X:Y]:=\sum_x\sum_yp_t(x,y)\ln\dfrac{p_t(x,y)}{p^X_t(x)p^Y_t(y)},
\end{align}
where $p^X_t(x)=\sum_yp_t(x,y)$ and $p^Y_t(y)=\sum_xp_t(x,y)$ denote the marginal distributions for $X$ and $Y$, respectively.
The mutual information is nonnegative and is equal to zero if and only if $X$ and $Y$ are independent.

The directional information from one variable to the other can be quantified by \textit{information flow}~\cite{horowitz2014thermodynamics}, which is defined as
\begin{align}
\dot{I}^X&:=\lim_{dt\rightarrow0^+}\dfrac{I[X_{t+dt}:Y_t]-I[X_t:Y_t]}{dt}\notag\\
&=\sum_x\sum_{x'}\sum_yw^y_{xx'}p_t(x',y)\ln\dfrac{p_t(y|x)}{p_t(y|x')},\\
\dot{I}^Y&:=\lim_{dt\rightarrow0^+}\dfrac{I[X_t:Y_{t+dt}]-I[X_t:Y_t]}{dt}\notag\\
&=\sum_x\sum_y\sum_{y'}w^{yy'}_xp_t(x,y')\ln\dfrac{p_t(x|y)}{p_t(x|y')},
\end{align}
where $p_t(y|x)=p_t(x,y)/p^X_t(x)$ and $p_t(x|y)=p_t(x,y)/p^Y_t(y)$ denote the conditional probabilities.
From the bipartite property, the sum of $\dot{I}^X$ and $\dot{I}^Y$ gives the time derivative of the mutual information~\cite{chetrite2019information}:
\begin{align}
d_tI[X:Y]=\dot{I}^X+\dot{I}^Y.
\end{align}
In the steady state condition, $\dot{I}^X$ and $\dot{I}^Y$ have opposite signs because $d_tI[X:Y]=0$.
If $\dot{I}^X>0$, the correlation between $X$ and $Y$ increases due to transitions in $X$.
In other words, $X$ gains information about $Y$.
If $\dot{I}^X<0$, in contrast, $X_{t+dt}$ is less correlated with $Y_t$ than $X_t$.
In this case, the information is destroyed or exploited by $X$.

\subsection{Second law of information thermodynamics}
Here, we formulate the second law of information thermodynamics.
To this end, we impose the local detailed balance condition to ensure that the system is thermodynamically consistent~\cite{seifert2012stochastic,maes2021local,peliti2021stochastic}.
Then, the entropy change in the environment due to transitions in $X$ and $Y$ is identified as
\begin{align}
\dot{S}_{\mathrm{env}}&=\sum_x\sum_{x'}\sum_yw^y_{xx'}p_t(x',y)\ln\dfrac{w^y_{xx'}}{w^y_{x'x}}\notag\\
&\quad+\sum_x\sum_y\sum_{y'}w^{yy'}_xp_t(x,y')\ln\dfrac{w^{yy'}_x}{w^{y'y}_x}\notag\\
&=:\dot{S}_{\mathrm{env}}^X+\dot{S}_{\mathrm{env}}^Y.
\end{align}
The average rate of the system entropy is identified as the time derivative of the system's Shannon entropy $S[X,Y]:=-\sum_{x,y}p_t(x,y)\ln p_t(x,y)$:
\begin{align}
d_tS[X,Y]&=\sum_x\sum_{x'}\sum_yw^y_{xx'}p_t(x',y)\ln\dfrac{p_t(x',y)}{p_t(x,y)}\notag\\
&\quad+\sum_x\sum_y\sum_{y'}w^{yy'}_xp_t(x,y')\ln\dfrac{p_t(x,y')}{p_t(x,y)}.
\end{align}
Then, the total entropy production rate $\dot{\sigma}$ is given by
\begin{align}
\dot{\sigma}=d_tS[X,Y]+\dot{S}_{\mathrm{env}}\ge0,
% &=\sum_x\sum_{x'}\sum_yw^y_{xx'}p_t(x',y)\ln\dfrac{w^y_{xx'}p_t(x',y)}{w^y_{x'x}p_t(x,y)}\notag\\
% &\quad+\sum_x\sum_y\sum_{y'}w^{yy'}_xp_t(x,y')\ln\dfrac{w^{yy'}_xp_t(x,y')}{w^{y'y}_xp_t(x,y)}\notag\\
% &=\dot{\sigma}^X+\dot{\sigma}^Y\ge0.
\end{align}
where the nonnegativity is proved by using $\ln a\le a-1$ ($a\ge0$).
The nonnegativity of the total entropy production rate is a manifestation of the second law of thermodynamics and is sometimes called the second law of stochastic thermodynamics~\cite{peliti2021stochastic}.

From the bipartite property, $\dot{\sigma}$ can be decomposed into two parts:
\begin{align}
\dot{\sigma}=\dot{\sigma}^X+\dot{\sigma}^Y.
\end{align}
Here, $\dot{\sigma}^X$ and $\dot{\sigma}^Y$ denote the partial entropy production rate due to transitions in $X$ and $Y$, respectively~\cite{shiraishi2015fluctuation}:
\begin{align}
\dot{\sigma}^X&:=\sum_x\sum_{x'}\sum_yw^y_{xx'}p_t(x',y)\ln\dfrac{w^y_{xx'}p_t(x',y)}{w^y_{x'x}p_t(x,y)}\notag\\
% \dot{\sigma}^Y&:=\sum_x\sum_y\sum_{y'}w^{yy'}_xp_t(x,y')\ln\dfrac{w^{yy'}_xp_t(x,y')}{w^{y'y}_xp_t(x,y)}.
% &=\sum_x\sum_{x'}\sum_yw^y_{xx'}p_t(x',y)\ln\dfrac{w^y_{xx'}p^X_t(x')}{w^y_{x'x}p^X_t(x)}\notag\\
% &\quad-\sum_x\sum_{x'}\sum_yw^y_{xx'}p_t(x',y)\ln\dfrac{p_t(y|x)}{p_t(y|x')}\notag\\
&=\dot{S}^X_{\mathrm{tot}}-\dot{I}^X,\label{partial entropy production rate X}\\
\dot{\sigma}^Y&:=\sum_x\sum_y\sum_{y'}w^{yy'}_xp_t(x,y')\ln\dfrac{w^{yy'}_xp_t(x,y')}{w^{y'y}_xp_t(x,y)}\notag\\
&=\dot{S}^Y_{\mathrm{tot}}-\dot{I}^Y,
\label{partial entropy production rate Y}
\end{align}
where $\dot{S}^Z_{\mathrm{tot}}$ ($Z=X,Y$) can be interpreted as the entropy production rate associated with $Z$, which consists of the time derivative of $Z$'s Shannon entropy $S[Z]:=-\sum_zp^Z_t(z)\ln p^Z_t(z)$ and the entropy change in the environment due to transitions in $Z$:
\begin{align}
\dot{S}^X_{\mathrm{tot}}&:=\sum_x\sum_{x'}\sum_yw^y_{xx'}p_t(x',y)\ln\dfrac{w^y_{xx'}p^X_t(x')}{w^y_{x'x}p^X_t(x)}\notag\\
% &=\sum_x\sum_{x'}\sum_yw^y_{xx'}p_t(x',y)\ln\dfrac{p^X_t(x')}{p^X_t(x)}\notag\\
% &\quad+\sum_x\sum_{x'}\sum_yw^y_{xx'}p_t(x',y)\ln\dfrac{w^y_{xx'}}{w^y_{x'x}}\notag\\
&=d_tS[X]+\dot{S}^X_{\mathrm{env}},\\
\dot{S}^Y_{\mathrm{tot}}&:=\sum_x\sum_y\sum_{y'}w^{yy'}_xp_t(x,y')\ln\dfrac{w^{yy'}_xp^Y_t(y')}{w^{y'y}_xp^Y_t(y)}\notag\\
&=d_tS[Y]+\dot{S}^Y_{\mathrm{env}}.
\end{align}
From the definition of the partial entropy production rates (\ref{partial entropy production rate X}) and (\ref{partial entropy production rate Y}), it immediately follows that $\dot{\sigma}^X$ and $\dot{\sigma}^Y$ are individually nonnegative:
\begin{figure}[t]
\includegraphics[width=5.5cm]{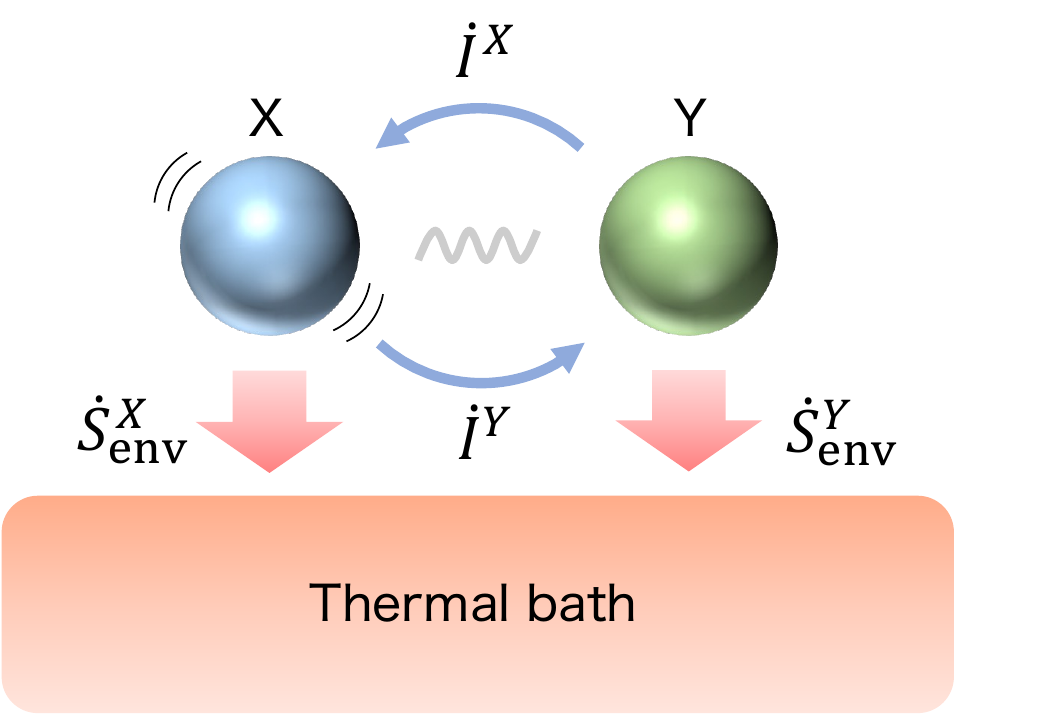}
\caption{Schematic of the second law of information thermodynamics. 
In the case where $X$ (blue) acts as a steady-state information-thermodynamic engine, $X$ converts information ($\dot{I}^X<0$) into negative entropy production ($\dot{S}^X_{\mathrm{env}}<0$), while $Y$ (green) gains information about $X$ ($\dot{I}^Y=-\dot{I}^X>0$) with the thermodynamic cost $\dot{S}^Y_{\mathrm{env}}>0$.
Although only a single thermal bath is depicted here, our results hold even in the presence of multiple thermal baths.}
\label{fig:second_law_information_thermodynamics}
\end{figure}
\begin{align}
\dot{\sigma}^X&=\dot{S}^X_{\mathrm{tot}}-\dot{I}^X\ge0,\label{second law of information thermodynamics_X}\\
\dot{\sigma}^Y&=\dot{S}^Y_{\mathrm{tot}}-\dot{I}^Y\ge0.
\end{align}
This is the so-called second law of information thermodynamics (see also Fig.~\ref{fig:second_law_information_thermodynamics}).
The important point here is that $\dot{S}^X_{\mathrm{tot}}$ ($\dot{S}^Y_{\mathrm{tot}}$) can be negative if $\dot{I}^X$ ($\dot{I}^Y$) is negative. 
This apparent violation of the second law of thermodynamics caused by information flow lies at the heart of the mechanism of Maxwell's demon~\cite{parrondo2015thermodynamics}.
In this case, $X$ acts as an information-thermodynamic engine that converts information into output work or negative entropy production.

\section{Thermodynamic uncertainty relation for bipartite systems\label{Thermodynamic uncertainty relation for bipartite systems}}
In this section, we explain our first main result, which can be regarded as an extension of the standard TUR (\ref{standard TUR}) to bipartite systems.
Hereafter, we assume that the whole system is in the steady state.
See Appendices~\ref{Bipartite TUR for overdamped Langevin equations} and \ref{Appendix: Transient TUR} for time-dependent cases.

\subsection{Bipartite TUR\label{Bipartite TUR}}
%%%%%% Bipartite TUR
Let $\hat{\mathcal{J}}_\mathcal{T}$ be a generalized time-integrated current for the subsystem $X$ with an arbitrary antisymmetric weight $d^y_{xx'}=-d^y_{x'x}$:
\begin{align}
\hat{\mathcal{J}}_\mathcal{T}:=\sum_{x}\sum_{x'(\neq x)}\sum_y\hat{n}^y_{xx'}d^y_{xx'},
\end{align}
where $\hat{n}^y_{xx'}$ denotes the number of transitions from the state $(x',y)$ to $(x,y)$ during the time interval $[0,\mathcal{T}]$.
For example, the choice of $d^y_{xx'}=\ln w^y_{xx'}/w^y_{x'x}$ yields the stochastic entropy change in the environment due to transitions in $X$ during $[0,\mathcal{T}]$.
In the steady state, the time-integrated stochastic information flow can also be expressed in this form (see (\ref{time-integrated stochastic information flow})).
We remark that when there are multiple environments with the label $\nu=1,2,\cdots$, the weight $d^y_{xx'}$ can depend on $\nu$.
The ensemble average of $\hat{\mathcal{J}}_\mathcal{T}$ reads
\begin{align}
\langle \hat{\mathcal{J}}_\mathcal{T}\rangle=\int^\mathcal{T}_0dt\sum_{x}\sum_{x'(\neq x)}\sum_yw^y_{xx'}p_{\mathrm{ss}}(x',y)d^y_{xx'}.
\end{align}

Our first main result is the following inequality:
for arbitrary observation time $\mathcal{T}$,
\begin{align}
\dfrac{\mathrm{Var}[\hat{\mathcal{J}}_\mathcal{T}]}{\langle \hat{\mathcal{J}}_\mathcal{T}\rangle^2}\ge\dfrac{2(1+\delta_{\mathcal{J}})^2}{\Delta S^X_{\mathrm{tot}}-\Delta I^X},
% \dfrac{\mathrm{Var}[\hat{\mathcal{J}}_\mathcal{T}]}{\left[\langle \hat{\mathcal{J}}_\mathcal{T}\rangle+\langle \hat{\mathcal{J}}_\mathcal{T}\rangle_q\right]^2}\ge\dfrac{2}{\Delta S^X_{\mathrm{tot}}-\Delta I^X},
\label{bipartite TUR}
\end{align}
where $\Delta S^X_{\mathrm{tot}}:=\int^\mathcal{T}_0dt\dot{S}^X_{\mathrm{tot}}$ and $\Delta I^X:=\int^\mathcal{T}_0dt\dot{I}^X$ denote the entropy production and time-integrated information flow associated with $X$, respectively.
Here, $\delta_{\mathcal{J}}:=\langle \hat{\mathcal{J}}_\mathcal{T}\rangle_q/\langle \hat{\mathcal{J}}_\mathcal{T}\rangle$, and $\langle \hat{\mathcal{J}}_\mathcal{T}\rangle_q$ is defined as
\begin{align}
\langle \hat{\mathcal{J}}_\mathcal{T}\rangle_q:=\int^\mathcal{T}_0dt\sum_{x}\sum_{x'(\neq x)}\sum_yw^y_{xx'}q_t(x',y)d^y_{xx'},
\end{align}
where $q_t$ satisfies the following equation with $q_0=0$:
\begin{align}
\partial_tq_t(x,y)&=\left[\sum_{x'}w^y_{xx'}q_t(x',y)+\sum_{y'}w^{yy'}_xq_t(x,y')\right]\notag\\
&\quad+\sum_{x'}w^y_{xx'}p_{\mathrm{ss}}(x',y).
\label{q-dynamics}
\end{align}

This additional current term $\langle \hat{\mathcal{J}}_\mathcal{T}\rangle_q$ reflects the contribution of the interaction with $Y$.
Indeed, if the transition rate for $X$ and the weight are independent of $Y$, i.e., $w^y_{xx'}=w_{xx'}$ and $d^y_{xx'}=d_{xx'}$, we can prove that $\langle \hat{\mathcal{J}}_\mathcal{T}\rangle_q=0$.
For the derivation, see Appendix~\ref{Appendix: Transient TUR}, where we consider the bipartite TUR in a more general case, applicable to a transient state.
If, in addition, $X$ and $Y$ are independent and thus $\Delta I^X=0$, then the bipartite TUR (\ref{bipartite TUR}) reduces to the standard TUR (\ref{standard TUR}), where the relative fluctuation of the current for $X$ is lower bounded by the entropy production associated with $X$.
In the general case where $X$ and $Y$ are correlated, the bipartite TUR (\ref{bipartite TUR}) states that the relative fluctuation of the current for the subsystem $X$ is lower bounded not only by the entropy production associated with $X$, but also by the information transfer between $X$ and $Y$.

There are two important cases where the additional current term $\langle \hat{\mathcal{J}}_\mathcal{T}\rangle_q$ can be ignored and thus $\delta_{\mathcal{J}}\rightarrow0$.
The first case occurs in the short time limit $\mathcal{T}\rightarrow0$.
Since $q_0=0$, it immediately follows that $\langle \hat{\mathcal{J}}_\mathcal{T}\rangle_q\rightarrow0$ as $\mathcal{T}\rightarrow0$, while the information flow $\Delta I^X$ remains finite in general.
The second case occurs in the long time limit $\mathcal{T}\rightarrow\infty$ where there is a separation of time scales between $X$ and $Y$.
That is, the case where the observation time is long and $Y$ evolves much faster than $X$.
The proof of $\delta_{\mathcal{J}}\rightarrow0$ in this case will be described in detail in Sec.~\ref{Fast relaxation limit for Y}.
Since this case is typically realized when $X$ acts as an information-thermodynamic engine, we will mainly focus on this case in the following sections.

\subsection{Derivation of the bipartite TUR\label{Derivation of the bipartite TUR}}
Here, we prove the bipartite TUR (\ref{bipartite TUR}) by using the generalized Cram\'er-Rao inequality~\cite{cover1999elements,dechant2018multidimensional,hasegawa2019uncertainty,shiraishi2021optimal}.
We remark that the bipartite TUR can also be proved more directly from the master equation or Langevin equation~\cite{dieball2023direct} (see also Appendix~\ref{Bipartite TUR for overdamped Langevin equations} for the direct derivation for overdamped Langevin equations).
We consider the following auxiliary dynamics parameterized by $\theta$ with $p^\theta_0=p_{\mathrm{ss}}$:
\begin{align}
\partial_tp^\theta_t(x,y)=\sum_{x'}w^y_{xx'}(\theta)p^\theta_t(x',y)+\sum_{y'}w^{yy'}_xp^\theta_t(x,y').
\label{auxiliary master equation}
\end{align}
Here, $w^y_{xx'}(\theta)$ denotes the parameterized transition rate
\begin{align}
w^y_{xx'}(\theta)&:=w^y_{xx'}e^{\theta Z^y_{xx'}}\qquad\text{for}\quad x\neq x',\notag\\
w^y_{xx}(\theta)&:=-\sum_{x'(\neq x)}w^y_{x'x}e^{\theta Z^y_{x'x}},
\end{align}
where
\begin{align}
Z^y_{xx'}:=\dfrac{w^y_{xx'}p_{\mathrm{ss}}(x',y)-w^y_{x'x}p_{\mathrm{ss}}(x,y)}{w^y_{xx'}p_{\mathrm{ss}}(x',y)+w^y_{x'x}p_{\mathrm{ss}}(x,y)}.
\end{align}
Let $\mathbb{P}_\theta(\Gamma)$ be the parameterized path probability for the trajectory $\Gamma=\{x_t,y_t\}^\mathcal{T}_{t=0}$,
\begin{widetext}
\begin{multline}
\mathbb{P}_\theta(\Gamma)=p_{\mathrm{ss}}(x_0,y_0)\exp\left[-\sum_x\sum_y\left(\sum_{x'(\neq x)}w^y_{x'x}(\theta)+\sum_{y'(\neq y)}w^{y'y}_x\right)\hat{\tau}^y_x\right.\\
\left.+\sum_x\sum_y\left(\sum_{x'(\neq x)}\hat{n}^y_{xx'}\ln w^y_{xx'}(\theta)+\sum_{y'(\neq y)}\hat{n}^{yy'}_x\ln w^{yy'}_x\right)\right],
\label{path probability}
\end{multline}
\end{widetext}
where $\hat{n}^{yy'}_x$ denotes the number of transitions from the state $(x,y')$ to $(x,y)$ during the time interval $[0,\mathcal{T}]$, and $\hat{\tau}^y_x$ denotes the empirical dwell time in state $(x,y)$, defined as the total amount of time spent in state $(x,y)$ along the trajectory $\Gamma$:
\begin{align}
\hat{\tau}^y_x:=\int^{\mathcal{T}}_0dt\delta_{xx_t}\delta^{yy_t},
\end{align}
where $\delta_{xx_t}$ ($\delta^{yy_t}$) denotes the Kronecker delta, which is $1$ if $x=x_t$ ($y=y_t$), and zero otherwise.
We denote by $\mathbb{I}(\theta):=-\langle\partial^2_\theta\ln\mathbb{P}_\theta(\Gamma)\rangle_\theta$ the corresponding Fisher information~\cite{cover1999elements}, where $\langle\cdot\rangle_\theta$ denotes the average with respect to $\mathbb{P}_\theta$.
The generalized Cram\'er-Rao inequality then yields~\cite{cover1999elements,dechant2018multidimensional,hasegawa2019uncertainty,shiraishi2021optimal}
\begin{align}
\dfrac{\mathrm{Var}[\hat{\mathcal{J}}_\mathcal{T}]}{\left(\partial_\theta\langle \hat{\mathcal{J}}_\mathcal{T}\rangle_\theta|_{\theta=0}\right)^2}\ge\dfrac{1}{\mathbb{I}(0)}.
\label{Cramer-Rao}
\end{align}
Here, $\mathbb{I}(0)$ can be expressed as
\begin{widetext}
\begin{align}
\mathbb{I}(0)&=\left.\left\langle\sum_x\sum_y\sum_{x'(\neq x)}\dfrac{\partial^2}{\partial\theta^2}w^y_{x'x}(\theta)\hat{\tau}^y_x-\sum_x\sum_y\sum_{x'(\neq x)}\dfrac{\partial^2}{\partial\theta^2}\ln w^y_{xx'}(\theta)\hat{n}^y_{xx'}\right\rangle_\theta\right|_{\theta=0}\notag\\
% &=\left.-\int^\mathcal{T}_0dt\left[\sum_x\sum_{x'(\neq x)}\sum_yw^y_{xx'}(\theta)p^\theta_t(x',y)\dfrac{\partial^2}{\partial\theta^2}\ln w^y_{xx'}(\theta)+\sum_x\sum_yp^\theta_t(x,y)\dfrac{\partial^2}{\partial\theta^2}\ln e^{w^y_{xx}(\theta)}\right]\right|_{\theta=0}\notag\\
&=\int^\mathcal{T}_0dt\sum_x\sum_{x'(\neq x)}\sum_yw^y_{xx'}p_{\mathrm{ss}}(x',y)\left(Z^y_{xx'}\right)^2\notag\\
&=\dfrac{1}{2}\int^\mathcal{T}_0dt\sum_x\sum_{x'(\neq x)}\sum_y\dfrac{\left[w^y_{xx'}p_{\mathrm{ss}}(x',y)-w^y_{x'x}p_{\mathrm{ss}}(x,y)\right]^2}{w^y_{xx'}p_{\mathrm{ss}}(x',y)+w^y_{x'x}p_{\mathrm{ss}}(x,y)}\notag\\
&\le\dfrac{1}{2}\int^\mathcal{T}_0dt\sum_x\sum_{x'(\neq x)}\sum_yw^y_{xx'}p_{\mathrm{ss}}(x',y)\ln\dfrac{w^y_{xx'}p_{\mathrm{ss}}(x',y)}{w^y_{x'x}p_{\mathrm{ss}}(x,y)}=\dfrac{\Delta S^X_{\mathrm{tot}}-\Delta I^X}{2},
\end{align}
\end{widetext}
where we have used the inequality $2(a-b)^2/(a+b)\le(a-b)\ln a/b$.
To calculate $\partial_\theta\langle \hat{\mathcal{J}}_\mathcal{T}\rangle_\theta|_{\theta=0}$, we expand $p^\theta_t(x,y)$ in terms of $\theta$ as $p^\theta_t(x,y)=p_{\mathrm{ss}}(x,y)+\theta q_t(x,y)+O(\theta^2)$.
By substituting this expression into (\ref{auxiliary master equation}), we find that
\begin{align}
\partial_tq_t(x,y)&=\left[\sum_{x'}w^y_{xx'}q_t(x',y)+\sum_{y'}w^{yy'}_xq_t(x,y')\right]\notag\\
&\quad+\sum_{x'}w^y_{xx'}p_{\mathrm{ss}}(x',y).
% \label{q-dynamics}
\end{align}
Then, $\partial_\theta\langle \hat{\mathcal{J}}_\mathcal{T}\rangle_\theta|_{\theta=0}$ can be calculated as
\begin{align}
&\partial_\theta\langle \hat{\mathcal{J}}_\mathcal{T}\rangle_\theta|_{\theta=0}\notag\\
=&\left.\partial_\theta\int^\mathcal{T}_0dt\sum_{x}\sum_{x'(\neq x)}\sum_yw^y_{xx'}(\theta)p^\theta_t(x',y)d^y_{xx'}\right|_{\theta=0}\notag\\
=&\langle \hat{\mathcal{J}}_\mathcal{T}\rangle+\langle \hat{\mathcal{J}}_\mathcal{T}\rangle_q.
\end{align}
We thus arrive at the inequality (\ref{bipartite TUR}).

\subsection{Fast relaxation limit of $Y$\label{Fast relaxation limit for Y}}
Here, we show that $\langle \hat{\mathcal{J}}_\mathcal{T}\rangle_q\ll\langle \hat{\mathcal{J}}_\mathcal{T}\rangle$ in the long time limit $\mathcal{T}\rightarrow\infty$ if $Y$ relaxes much faster than $X$.
Let $\tau_X$ and $\tau_Y$ be the time scale of $X$ and $Y$, respectively.
We assume a separation of time scales: $\tau_Y\ll\tau_X$, i.e., the auxiliary system $Y$ evolves much faster than the system of interest $X$.
This situation is typically realized when $Y$ acts as a Maxwell's demon, i.e., when $Y$ measures the state of $X$ and performs feedback control~\cite{horowitz2014thermodynamics}.
We introduce a dimensionless slow time $\tau:=t/\tau_X$ and a small parameter $\epsilon:=\tau_Y/\tau_X\ll1$.
Correspondingly, we nondimensionalize the transition rates as $\widetilde{w}^y_{xx'}:=\tau_Xw^y_{xx'}$ and $\widetilde{w}^{yy'}_x:=\tau_Yw^{yy'}_x$.

We first take the long time limit $\mathcal{T}\rightarrow\infty$, i.e., $\mathcal{T}\gg\tau_X$, and assume that $q_t(x,y)$ reaches a stationary solution $q_{\mathrm{ss}}(x,y)$.
Then, $p_{\mathrm{ss}}$ and $q_{\mathrm{ss}}$ satisfy the following equations:
\begin{align}
\sum_{x'}\widetilde{w}^y_{xx'}p_{\mathrm{ss}}(x',y)+\dfrac{1}{\epsilon}\sum_{y'}\widetilde{w}^{yy'}_xp_{\mathrm{ss}}(x,y')=0,\label{dimensionless master equation}
% \partial_\tau p_\tau(x,y)&=\sum_{x'}\widetilde{w}^y_{xx'}p_\tau(x',y)+\dfrac{1}{\epsilon}\sum_{y'}\widetilde{w}^{yy'}_xp_\tau(x,y'),\label{dimensionless master equation}\\
\end{align}
\begin{multline}
\left[\sum_{x'}\widetilde{w}^y_{xx'}q_{\mathrm{ss}}(x',y)+\dfrac{1}{\epsilon}\sum_{y'}\widetilde{w}^{yy'}_xq_{\mathrm{ss}}(x,y')\right]\\
+\sum_{x'}\widetilde{w}^y_{xx'}p_{\mathrm{ss}}(x',y)=0.\label{dimensionless q-dynamics}
% \partial_\tau q_\tau(x,y)&=\left[\sum_{x'}\widetilde{w}^y_{xx'}q_\tau(x',y)+\dfrac{1}{\epsilon}\sum_{y'}\widetilde{w}^{yy'}_xq_\tau(x,y')\right]\notag\\
% &\quad+\sum_{x'}\widetilde{w}^y_{xx'}p_\tau(x',y).\label{dimensionless q-dynamics}
\end{multline}
We now assume that $p_{\mathrm{ss}}$ and $q_{\mathrm{ss}}$ have asymptotic expansions in terms of the asymptotic sequences $\{\epsilon^n\}^\infty_{n=0}$ as $\epsilon\rightarrow0$:
\begin{align}
p_{\mathrm{ss}}&=p^{(0)}_{\mathrm{ss}}+\epsilon p^{(1)}_{\mathrm{ss}}+\cdots,\\
q_{\mathrm{ss}}&=q^{(0)}_{\mathrm{ss}}+\epsilon q^{(1)}_{\mathrm{ss}}+\cdots.
\end{align}
Here, we impose the normalization condition
\begin{align}
\sum_yp^{(0)}_{\mathrm{ss}}(x,y)&=p^X_{\mathrm{ss}}(x),\\
\sum_yq^{(0)}_{\mathrm{ss}}(x,y)&=q^X_{\mathrm{ss}}(x),
\end{align}
where we have introduced $q^X_{\mathrm{ss}}(x):=\sum_yq_{\mathrm{ss}}(x,y)$.
Note that $q^X_{\mathrm{ss}}$ satisfies the normalization condition $\sum_xq^X_{\mathrm{ss}}(x)=0$.

By substituting these expansions into (\ref{dimensionless master equation}) and (\ref{dimensionless q-dynamics}), we find that the leading order yields
\begin{align}
\sum_{y'}\widetilde{w}^{yy'}_xp^{(0)}_{\mathrm{ss}}(x,y')&=0,\\
\sum_{y'}\widetilde{w}^{yy'}_xq^{(0)}_{\mathrm{ss}}(x,y')&=0.
\end{align}
Let $\pi_{\mathrm{ss}}(y|x)$ be the normalized zero-eigenvector that satisfies $\sum_{y'}\widetilde{w}^{yy'}_x\pi_{\mathrm{ss}}(y'|x)=0$.
Due to the irreducibility of the rate matrix, this normalized zero-eigenvector is unique for each $x$.
Then, from the normalization condition, $p^{(0)}_{\mathrm{ss}}$ and $q^{(0)}_{\mathrm{ss}}$ should have the form
\begin{align}
p^{(0)}_{\mathrm{ss}}(x,y)&=p^X_{\mathrm{ss}}(x)\pi_{\mathrm{ss}}(y|x),\\
q^{(0)}_{\mathrm{ss}}(x,y)&=q^X_{\mathrm{ss}}(x)\pi_{\mathrm{ss}}(y|x).
\end{align}

The subleading order of (\ref{dimensionless master equation}) and (\ref{dimensionless q-dynamics}) yields
\begin{align}
\sum_{x'}\widetilde{w}^y_{xx'}p^{(0)}_{\mathrm{ss}}(x',y)+\sum_{y'}\widetilde{w}^{yy'}_xp^{(1)}_{\mathrm{ss}}(x,y')=0,
% \partial_\tau p^{(0)}_\tau(x,y)&=\sum_{x'}\widetilde{w}^y_{xx'}p^{(0)}_\tau(x',y)+\sum_{y'}\widetilde{w}^{yy'}_xp^{(1)}_\tau(x,y'),
\label{subleading order_p}
\end{align}
\begin{multline}
\left[\sum_{x'}\widetilde{w}^y_{xx'}q^{(0)}_{\mathrm{ss}}(x',y)+\sum_{y'}\widetilde{w}^{yy'}_xq^{(1)}_{\mathrm{ss}}(x,y')\right]\\
+\sum_{x'}\widetilde{w}^y_{xx'}p^{(0)}_{\mathrm{ss}}(x',y)=0.
% \partial_\tau q^{(0)}_\tau(x,y)&=\left[\sum_{x'}\widetilde{w}^y_{xx'}q^{(0)}_\tau(x',y)+\sum_{y'}\widetilde{w}^{yy'}_xq^{(1)}_\tau(x,y')\right]\notag\\
% &\quad+\sum_{x'}\widetilde{w}^y_{xx'}p^{(0)}_\tau(x',y).
\label{subleading order_q}
\end{multline}
Note that (\ref{subleading order_p}) and (\ref{subleading order_q}) are linear equations for $p^{(1)}_{\mathrm{ss}}$ and $q^{(1)}_{\mathrm{ss}}$ with the matrix $\widetilde{w}^{yy'}_x$, which has the left zero-eigenvector $1$ because $\sum_y\widetilde{w}^{yy'}_x=0$.
This property guarantees that the solutions $p^{(1)}_{\mathrm{ss}}$ and $q^{(1)}_{\mathrm{ss}}$ exist only under the solvability conditions:
\begin{align}
\sum_{x'}\overline{w}_{xx'}p^X_{\mathrm{ss}}(x')&=0,\\
\sum_{x'}\overline{w}_{xx'}q^X_{\mathrm{ss}}(x')&=0,\label{q^X dynamics}
% \partial_\tau p^X_\tau(x)&=\sum_{x'}\overline{w}_{xx'}p^X_\tau(x'),\\
% \partial_\tau q^X_\tau(x)&=\sum_{x'}\overline{w}_{xx'}q^X_\tau(x')+\sum_{x'}\overline{w}_{xx'}p^X_\tau(x'),\label{q^X dynamics}
\end{align}
which correspond to (\ref{subleading order_p}) and (\ref{subleading order_q}) summed over $y$, respectively.
Here, we have introduced the effective transition rate $\overline{w}_{xx'}:=\sum_y\widetilde{w}^{y}_{xx'}\pi_{\mathrm{ss}}(y|x')$.
Then, from the Perron-Frobenius theorem, $q^X_{\mathrm{ss}}$ can be expressed as $q^X_{\mathrm{ss}}=\mathcal{N}p^X_{\mathrm{ss}}$, where $\mathcal{N}$ denotes the normalization constant.
Because $q^X_{\mathrm{ss}}$ satisfies the normalization condition $\sum_xq^X_{\mathrm{ss}}(x)=0$, we obtain $\mathcal{N}=0$.
Thus, in the fast relaxation limit of $Y$, we have
\begin{align}
p_{\mathrm{ss}}(x,y)&=p^X_{\mathrm{ss}}(x)\pi_{\mathrm{ss}}(y|x)+O(\epsilon),\\
q_{\mathrm{ss}}(x,y)&=O(\epsilon).
\end{align}
Therefore, the additional current term $\langle \hat{\mathcal{J}}_\mathcal{T}\rangle_q$ appearing in the bipartite TUR (\ref{bipartite TUR}) is much smaller than $\langle \hat{\mathcal{J}}_\mathcal{T}\rangle$ in the long time limit $\mathcal{T}\rightarrow\infty$:
\begin{align}
% \langle \hat{\mathcal{J}}_\mathcal{T}\rangle_q\ll\langle \hat{\mathcal{J}}_\mathcal{T}\rangle.
\delta_{\mathcal{J}}:=\dfrac{\langle \hat{\mathcal{J}}_\mathcal{T}\rangle_q}{\langle \hat{\mathcal{J}}_\mathcal{T}\rangle}\rightarrow0.
\end{align}

To summarize, in the fast relaxation limit of $Y$, the bipartite TUR (\ref{bipartite TUR}) reduces to the form similar to the standard TUR (\ref{standard TUR}) in the long time limit:
\begin{align}
\dfrac{D_{\mathcal{J}}}{J^2}\ge\dfrac{1}{\dot{S}^X_{\mathrm{tot}}-\dot{I}^X},
\label{bipartite TUR_fast relaxation limit}
\end{align}
where $D_{\mathcal{J}}:=\lim_{\mathcal{T}\rightarrow\infty}\mathrm{Var}[\hat{\mathcal{J}}_{\mathcal{T}}]/2\mathcal{T}$ denotes the fluctuation of $\hat{\mathcal{J}}_{\mathcal{T}}$, and $J:=\lim_{\mathcal{T}\rightarrow\infty}\langle\hat{\mathcal{J}}_{\mathcal{T}}\rangle/\mathcal{T}$ denotes the mean current.
Note that (\ref{bipartite TUR_fast relaxation limit}) gives a tighter lower bound on the fluctuation of a current than the standard TUR, because $\dot{S}^X_{\mathrm{tot}}-\dot{I}^X$ is smaller than or equal to the total entropy production $\dot{\sigma}$.
If the partial entropy production of $Y$ is zero, then (\ref{bipartite TUR_fast relaxation limit}) can also be obtained from the standard TUR.

%%% Equality condition
\subsection{Equality condition\label{Equality condition}}
The equality of the bipartite TUR in the fast relaxation limit of $Y$ (\ref{bipartite TUR_fast relaxation limit}) can be achieved even far from equilibrium.
This nontrivial fact will be shown later with a simple example in Sec.~\ref{Coupled linear overdamped Langevin equations}. 
This property is a stark difference from the standard TUR, where the equality is guaranteed only in the near-equilibrium limit~\cite{hasegawa2019uncertainty,shiraishi2021optimal}.
Here, before showing the example in Sec.~\ref{Coupled linear overdamped Langevin equations}, we discuss a possible scenario to achieve the equality of the bipartite TUR (\ref{bipartite TUR}) in a somewhat abstract manner.

We first consider the equality condition of the generalized Cram\'er-Rao inequality at $\theta=0$ (\ref{Cramer-Rao}).
Because the generalized Cram\'er-Rao inequality is based on the Cauchy-Schwarz inequality, the equality condition is satisfied if and only if the following relation holds ~\cite{hasegawa2019uncertainty}:
\begin{align}
\hat{\mathcal{J}}_{\mathcal{T}}-\langle\hat{\mathcal{J}}_{\mathcal{T}}\rangle=C\left.\dfrac{\partial}{\partial\theta}\ln\mathbb{P}_\theta(\Gamma)\right|_{\theta=0},
\label{equality condition}
\end{align}
where $C$ is a constant.
The right-hand side of (\ref{equality condition}) is given by
\begin{align}
C\left.\dfrac{\partial}{\partial\theta}\ln\mathbb{P}_\theta(\Gamma)\right|_{\theta=0}=C\sum_x\sum_{x'(\neq x)}\sum_yZ^y_{xx'}(\hat{n}^y_{xx'}-w^y_{xx'}\hat{\tau}^y_{x'}).
\label{LHS_equality condition}
\end{align}
The left-hand side of (\ref{equality condition}) reads
\begin{align}
\hat{\mathcal{J}}_{\mathcal{T}}-\langle\hat{\mathcal{J}}_{\mathcal{T}}\rangle&=\hat{\mathcal{J}}^{\mathrm{I}}_{\mathcal{T}}+\hat{\mathcal{J}}^{\mathrm{II}}_{\mathcal{T}}-\langle\hat{\mathcal{J}}_{\mathcal{T}}\rangle,
\label{RHS_equality condition}
\end{align}
where we have decomposed the current $\hat{\mathcal{J}}_{\mathcal{T}}$ into two parts~\cite{dieball2023direct}:
\begin{align}
\hat{\mathcal{J}}^{\mathrm{I}}_{\mathcal{T}}&:=\sum_{x}\sum_{x'(\neq x)}\sum_yd^y_{xx'}(\hat{n}^y_{xx'}-w^y_{xx'}\hat{\tau}^y_{x'}),\\
\hat{\mathcal{J}}^{\mathrm{II}}_{\mathcal{T}}&:=\sum_{x}\sum_{x'(\neq x)}\sum_yd^y_{xx'}w^y_{xx'}\hat{\tau}^y_{x'}.
\end{align}
Note that $\langle\hat{\mathcal{J}}^{\mathrm{I}}_{\mathcal{T}}\rangle=0$ and $\langle\hat{\mathcal{J}}^{\mathrm{II}}_{\mathcal{T}}\rangle=\langle\hat{\mathcal{J}}_{\mathcal{T}}\rangle$.
By comparing (\ref{LHS_equality condition}) and (\ref{RHS_equality condition}), we expect that $d^y_{xx'}=CZ^y_{xx'}$ to be the optimal choice.
However, due to the presence of $\hat{\mathcal{J}}^{\mathrm{II}}_{\mathcal{T}}-\langle\hat{\mathcal{J}}_{\mathcal{T}}\rangle$, the equality condition is generally not satisfied.

In the standard TUR, it is known that the equality can be achieved by including the generalized time-integrated static observable in addition to the current $\hat{\mathcal{J}}_{\mathcal{T}}$~\cite{dechant2021improving,shiraishi2021optimal,dieball2023direct}.
Here, we consider the following generalized time-integrated static observable:
\begin{align}
\hat{\mathcal{O}}_{\mathcal{T}}:=\sum_{x}\sum_y\rho^y_x\hat{\tau}^y_x,
\end{align}
where $\rho^y_x$ is an arbitrary weight that depends on a state $(x,y)$.
Even for the observable $\hat{\mathcal{J}}_{\mathcal{T}}+\hat{\mathcal{O}}_{\mathcal{T}}$ instead of $\hat{\mathcal{J}}_{\mathcal{T}}$, by following the same argument described in Sec.~\ref{Derivation of the bipartite TUR}, we can derive the following \textit{bipartite-correlation} TUR:
\begin{align}
\dfrac{\mathrm{Var}[\hat{\mathcal{J}}_\mathcal{T}+\hat{\mathcal{O}}_\mathcal{T}]}{\left[\langle \hat{\mathcal{J}}_\mathcal{T}\rangle+\langle \hat{\mathcal{J}}_\mathcal{T}\rangle_q+\langle \hat{\mathcal{O}}_\mathcal{T}\rangle_q\right]^2}&\ge\mathbb{I}(0)\label{bipartite correlation TUR_Cauchy-Schwarz}\\
&\ge\dfrac{2}{\Delta S^X_{\mathrm{tot}}-\Delta I^X},
\label{bipartite correlation TUR}
\end{align}
where $\langle \hat{\mathcal{O}}_\mathcal{T}\rangle_q$ is defined as
\begin{align}
\langle \hat{\mathcal{O}}_\mathcal{T}\rangle_q:=\int^\mathcal{T}_0dt\sum_{x}\sum_yq_t(x,y)\rho^y_x.
\end{align}
In this case, the equality condition of the inequality (\ref{bipartite correlation TUR_Cauchy-Schwarz}) is given by
\begin{align}
\hat{\mathcal{J}}_{\mathcal{T}}+\hat{\mathcal{O}}_{\mathcal{T}}-\langle\hat{\mathcal{J}}_{\mathcal{T}}+\hat{\mathcal{O}}_{\mathcal{T}}\rangle=C\left.\dfrac{\partial}{\partial\theta}\ln\mathbb{P}_\theta(\Gamma)\right|_{\theta=0}.
\label{equality condition_correlation TUR}
\end{align}
Then, we find that the choice
\begin{align}
d^y_{xx'}&=CZ^y_{xx'},\label{choice_d}\\
\rho^y_x&=-\sum_{x'(\neq x)}d^y_{x'x}w^y_{x'x},\label{choice_rho}
\end{align}
yields
\begin{align}
\hat{\mathcal{J}}_{\mathcal{T}}+\hat{\mathcal{O}}_{\mathcal{T}}-\langle\hat{\mathcal{J}}_{\mathcal{T}}+\hat{\mathcal{O}}_{\mathcal{T}}\rangle=\hat{\mathcal{J}}^{\mathrm{I}}_{\mathcal{T}}=C\left.\dfrac{\partial}{\partial\theta}\ln\mathbb{P}_\theta(\Gamma)\right|_{\theta=0},
\end{align}
where we have used the fact that $\hat{\mathcal{O}}_{\mathcal{T}}=- \hat{\mathcal{J}}^{\mathrm{II}}_\mathcal{T}$ for this choice.
Thus, the equality of (\ref{bipartite correlation TUR_Cauchy-Schwarz}) is achieved for this choice of $\hat{\mathcal{J}}_{\mathcal{T}}$ and $\hat{\mathcal{O}}_{\mathcal{T}}$.
However, even if we can achieve the equality of (\ref{bipartite correlation TUR_Cauchy-Schwarz}), the equality condition of the second inequality (\ref{bipartite correlation TUR}) is not satisfied in general.
Still, in the overdamped Langevin case, the inequality $\mathbb{I}(0)\ge2/(\Delta S^X_{\mathrm{tot}}-\Delta I^X)$ becomes an equality (see~\cite{otsubo2020estimating} for the detailed discussion).
Therefore, the equality of the bipartite-correlation TUR is achieved in the overdamped Langevin case even far from equilibrium for the choice (\ref{choice_d}) and (\ref{choice_rho}):
\begin{align}
\dfrac{\mathrm{Var}[\hat{\mathcal{J}}^{\mathrm{I}}_\mathcal{T}]}{\langle \hat{\mathcal{J}}_\mathcal{T}\rangle^2}=\dfrac{2}{\Delta S^X_{\mathrm{tot}}-\Delta I^X},
\label{bipartite correlation TUR_equality}
\end{align}
where we have used $\langle\hat{\mathcal{O}}_{\mathcal{T}}\rangle_q=-\langle \hat{\mathcal{J}}_\mathcal{T}\rangle_q$.
In the long time limit $\mathcal{T}\rightarrow\infty$, (\ref{bipartite correlation TUR_equality}) can be rewritten as
\begin{align}
\dfrac{D^{\mathrm{I}}_{\mathcal{J}}}{J^2}=\dfrac{1}{\dot{S}^X_{\mathrm{tot}}-\dot{I}^X},
% \dfrac{\mathrm{Var}[\hat{\mathcal{J}}^{\mathrm{I}}_\mathcal{T}]}{\langle \hat{\mathcal{J}}_\mathcal{T}\rangle^2}=\dfrac{2}{\Delta S^X_{\mathrm{tot}}-\Delta I^X}.
\label{equality_bipartite correlation TUR_fast relaxation limit}
\end{align}
where $D^{\mathrm{I}}_{\mathcal{J}}:=\lim_{\mathcal{T}\rightarrow\infty}\mathrm{Var}[\hat{\mathcal{J}}^{\mathrm{I}}_{\mathcal{T}}]/2\mathcal{T}$ denotes the fluctuation of $\hat{\mathcal{J}}^{\mathrm{I}}_\mathcal{T}$.
Note that $D^{\mathrm{I}}_{\mathcal{J}}$ is generally different from $D_{\mathcal{J}}$, and thus the equality (\ref{equality_bipartite correlation TUR_fast relaxation limit}) does generally not correspond to the equality of the bipartite TUR in the fast relaxation limit of $Y$ (\ref{bipartite TUR_fast relaxation limit}).
To put it another way, if $D^{\mathrm{I}}_{\mathcal{J}}=D_{\mathcal{J}}$ in the fast relaxation limit, then the equality of (\ref{bipartite TUR_fast relaxation limit}) is achieved.
Since $\langle\hat{\mathcal{J}}^{\mathrm{I}}_\mathcal{T}\rangle=0$, the covariance between $\hat{\mathcal{J}}^{\mathrm{I}}_\mathcal{T}$ and $\hat{\mathcal{J}}^{\mathrm{II}}_\mathcal{T}$ reads
\begin{align}
\mathrm{Cov}[\hat{\mathcal{J}}^{\mathrm{I}}_\mathcal{T},\hat{\mathcal{J}}^{\mathrm{II}}_\mathcal{T}]&=\langle\hat{\mathcal{J}}^{\mathrm{I}}_\mathcal{T}\hat{\mathcal{J}}^{\mathrm{II}}_\mathcal{T}\rangle\notag\\
&=C\left\langle\left.\dfrac{\partial}{\partial\theta}\ln\mathbb{P}_\theta(\Gamma)\right|_{\theta=0}\hat{\mathcal{J}}^{\mathrm{II}}_\mathcal{T}\right\rangle\notag\\
&=C\langle \hat{\mathcal{J}}_\mathcal{T}\rangle_q.
\end{align}
Therefore, the difference between $D^{\mathrm{I}}_{\mathcal{J}}$ and $D_{\mathcal{J}}$ is given by
\begin{align}
D_{\mathcal{J}}-D^{\mathrm{I}}_{\mathcal{J}}&=\lim_{\mathcal{T}\rightarrow\infty}\dfrac{1}{2\mathcal{T}}\mathrm{Var}[\hat{\mathcal{J}}^{\mathrm{I}}_{\mathcal{T}}+\hat{\mathcal{J}}^{\mathrm{II}}_{\mathcal{T}}]-\lim_{\mathcal{T}\rightarrow\infty}\dfrac{1}{2\mathcal{T}}\mathrm{Var}[\hat{\mathcal{J}}^{\mathrm{I}}_{\mathcal{T}}]\notag\\
&=D^{\mathrm{II}}_{\mathcal{J}}+\lim_{\mathcal{T}\rightarrow\infty}\dfrac{C}{\mathcal{T}}\langle \hat{\mathcal{J}}_\mathcal{T}\rangle_q\notag\\
&=D^{\mathrm{II}}_{\mathcal{J}},
% \mathrm{Var}[\hat{\mathcal{J}}_\mathcal{T}]-\mathrm{Var}[\hat{\mathcal{J}}^{\mathrm{I}}_\mathcal{T}]\simeq\mathrm{Var}[\hat{\mathcal{J}}^{\mathrm{II}}_\mathcal{T}]
\label{Var_J-Var_J^I=Var_J^II}
\end{align}
in the fast relaxation limit of $Y$.
Thus, $D^{\mathrm{II}}_{\mathcal{J}}=0$ is a sufficient condition for (\ref{bipartite TUR_fast relaxation limit}) to hold with equality.
In Sec.~\ref{Coupled linear overdamped Langevin equations}, we give an example that satisfies this sufficient condition.

\section{Trade-off relations\label{Trade-off relations}}
In this section, we focus on the regime where $Y$ evolves much faster than $X$ and show that the bipartite TUR in this regime (\ref{bipartite TUR_fast relaxation limit}) provides trade-off relations for the performance of information processing systems.
In Sec.~\ref{Trade-offs for information-thermodynamic engines}, we consider the situation where the subsystem $X$ can be regarded as a steady-state information-thermodynamic engine, while the external system $Y$ plays the role of a memory of Maxwell's demon.
From the second law of information thermodynamics (\ref{second law of information thermodynamics_X}), this situation corresponds to the case where $0<-\dot{S}^X_{\mathrm{env}}\le-\dot{I}^X$.
While this situation may be typical in the regime of the fast relaxation limit of $Y$, we can also consider the case where the slow system $X$ plays the role of a memory and measures the state of the fast system $Y$, i.e., $0< \dot{I}^X \le \dot{S}^X_{\mathrm{env}}$.
Even for this case, we can show that the bipartite TUR (\ref{bipartite TUR_fast relaxation limit}) provides trade-off relations on the performance of the memory, which will be described in Sec.~\ref{In the case of positive information flow}.

\subsection{Information-thermodynamic engine: \protect \\
$0<-\dot{S}^X_{\mathrm{env}}\le-\dot{I}^X$ \label{Trade-offs for information-thermodynamic engines}}
Here, we show that the bipartite TUR gives several universal bounds on the performance of information-thermodynamic engines.
In this case, both the entropy production and the information flow associated with $X$ are negative and satisfy the relation $0<-\dot{S}^X_{\mathrm{env}}\le-\dot{I}^X$.
Then, the performance of an information-thermodynamic engine can be quantified by, e.g., the information-thermodynamic efficiency~\cite{horowitz2014thermodynamics}:
\begin{align}
\eta^X_S:=\dfrac{|\dot{S}^X_{\mathrm{env}}|}{|\dot{I}^X|},
\end{align}
which satisfies $0\le\eta^X_S\le1$ as a direct consequence of the second law of information thermodynamics.
This efficiency quantifies how efficiently the engine $X$ converts information into negative entropy production.
In addition to this information-thermodynamic efficiency, the negative entropy production rate itself is an important indicator characterizing the performance of an information-thermodynamic engine.
Here, we show that there is the following trade-off relation between $\eta^X_S$ and $|\dot{S}^X_{\mathrm{env}}|$:
\begin{align}
|\dot{S}^X_{\mathrm{env}}|\le D_S\dfrac{1-\eta^X_S}{\eta^X_S},
\label{trade-off between entropy production and efficiency}
\end{align}
where $D_S$ denotes the fluctuation of the stochastic medium entropy production $\Delta\hat{S}^X_{\mathrm{env}}$,
\begin{align}
D_S:=\lim_{\mathcal{T}\rightarrow\infty}\dfrac{1}{2\mathcal{T}}\mathrm{Var}[\Delta\hat{S}^X_{\mathrm{env}}].
\end{align}
This inequality states that an information engine with a finite negative entropy production rate cannot achieve $\eta^X_S=1$ as long as the fluctuation $D_S$ is finite.
In order to achieve a finite negative entropy production rate with $\eta^X_S=1$, the fluctuation $D_S$ must diverge.
We can also prove a similar trade-off relation where the negative entropy production rate is bounded by the fluctuation of the time-integrated stochastic information flow $\Delta\hat{I}^X$ instead of $D_S$:
\begin{align}
|\dot{S}^X_{\mathrm{env}}|\le D_I\eta^X_S(1-\eta^X_S),
\label{trade-off between entropy production and efficiency via D_I}
\end{align}
where
\begin{align}
D_I:=\lim_{\mathcal{T}\rightarrow\infty}\dfrac{1}{2\mathcal{T}}\mathrm{Var}[\Delta\hat{I}^X].
\end{align}
The inequalities (\ref{trade-off between entropy production and efficiency}) and (\ref{trade-off between entropy production and efficiency via D_I}) are the second main results of this paper.

In Sec.~\ref{Derivation of the trade-offs}, we provide detailed proof of these inequalities.
In Sec.~\ref{Tightness of the bounds}, we briefly discuss which of the two inequalities (\ref{trade-off between entropy production and efficiency}) and (\ref{trade-off between entropy production and efficiency via D_I}) gives a tighter bound on the negative entropy production rate.
In Sec.~\ref{Trade-off between power and efficiency}, we derive a trade-off relation in terms of power, i.e., output work produced per unit of time, instead of the negative entropy production.
This relation can be regarded as a direct extension of the trade-offs for heat engines~\cite{shiraishi2016universal,pietzonka2018universal} to information-thermodynamic engines. 
\\

\subsubsection{Derivation of (\ref{trade-off between entropy production and efficiency}) and (\ref{trade-off between entropy production and efficiency via D_I})\label{Derivation of the trade-offs}}
Here, we derive the trade-off relations (\ref{trade-off between entropy production and efficiency}) and (\ref{trade-off between entropy production and efficiency via D_I}) by using the bipartite TUR in the fast relaxation limit of $Y$ (\ref{bipartite TUR_fast relaxation limit}), which can be rewritten as follows:
\begin{align}
|J|\le D_{\mathcal{J}}\dfrac{\dot{S}^X_{\mathrm{env}}-\dot{I}^X}{|J|}.
%|J|\le D_{\mathcal{J}}\dfrac{\left(\dot{S}^X_{\mathrm{env}}-\dot{I}^X\right)}{|J|}.  
% \dfrac{1}{\mathcal{T}}|\langle \hat{\mathcal{J}}_\mathcal{T}\rangle|\le\dfrac{\mathrm{Var}[\hat{\mathcal{J}}_\mathcal{T}]}{2\mathcal{T}}\dfrac{\left(\dot{S}^X_{\mathrm{env}}-\dot{I}^X\right)}{|\langle \hat{\mathcal{J}}_\mathcal{T}\rangle|/\mathcal{T}}.
\label{Trade-off TUR}
\end{align}

Let us choose stochastic medium entropy production associated with $X$ as current $\hat{\mathcal{J}}_\mathcal{T}$ in (\ref{Trade-off TUR}).
\begin{align}
\hat{\mathcal{J}}_\mathcal{T}&=\Delta\hat{S}^X_{\mathrm{env}}\notag\\
&:=\sum_{x}\sum_{x'(\neq x)}\sum_y\hat{n}^y_{xx'}\ln\dfrac{w^y_{xx'}}{w^y_{x'x}},
\end{align} 
which satisfies $\langle\Delta\hat{S}^X_{\mathrm{env}}\rangle=\Delta S^X_{\mathrm{env}}$.
Then, we immediately obtain the trade-off between entropy production and efficiency (\ref{trade-off between entropy production and efficiency}).

Another type of the trade-off relation (\ref{trade-off between entropy production and efficiency via D_I}) can be derived by choosing the time-integrated stochastic information flow as current $\hat{\mathcal{J}}_\mathcal{T}$ in (\ref{Trade-off TUR}).
Here, the instantaneous stochastic information flow is defined as the partial rate of change of the stochastic mutual information $\hat{I}(x_t:y_t):=\ln[p_t(x_t,y_t)/p^X_t(x_t)p^Y_t(y_t)]$:
\begin{align}
\hat{\dot{I}}^X&:=\sum_n\delta(t-t_n)\ln\dfrac{p_{t_n}(y_{t_n}|x_{t^+_n})}{p_{t_n}(y_{t_n}|x_{t^-_n})}\notag\\
&\qquad+\left.\dfrac{1}{p_t(x,y)}\sum_{x'}w^y_{xx'}p_t(x',y)\right|_{(x,y)=(x_t,y_t)}\notag\\
&\qquad-\left.\dfrac{1}{p^X_t(x)}\sum_{x'}\overline{w}_{xx'}p^X_t(x')\right|_{x=x_t},
\end{align}
where $t_n$ denotes the time at which $X$ jumps from $x_{t^-_n}$ to $x_{t^+_n}$, and $\overline{w}_{xx'}:=\sum_yw^y_{xx'}p_t(y|x')$ denotes the effective transition rate.
In the steady state, the last two terms vanish so that the time-integrated stochastic information flow reads
\begin{align}
\Delta\hat{I}^X=\sum_{x}\sum_{x'(\neq x)}\sum_y\hat{n}^y_{xx'}\ln\dfrac{\pi_{\mathrm{ss}}(y|x)}{\pi_{\mathrm{ss}}(y|x')},
% &:=\sum_{x}\sum_{x'(\neq x)}\sum_y\hat{n}^y_{xx'}\ln\dfrac{p_t(y|x)}{p_t(y|x')}.
\label{time-integrated stochastic information flow}
\end{align}
which satisfies $\langle\Delta\hat{I}^X\rangle=\Delta I^X$.
Substituting $\hat{\mathcal{J}}_{\mathcal{T}}=\Delta\hat{I}^X$ in the bipartite TUR (\ref{Trade-off TUR}), we obtain the inequality (\ref{trade-off between entropy production and efficiency via D_I}).
Note that we can also obtain a trade-off relation between the information flow and efficiency:
\begin{align}
|\dot{I}^X|\le D_I(1-\eta^X_S).
\label{trade-off between information and efficiency}
\end{align}

\subsubsection{Tightness of the bounds\label{Tightness of the bounds}}
Here, we consider which of the two inequalities (\ref{trade-off between entropy production and efficiency}) and (\ref{trade-off between entropy production and efficiency via D_I}) gives a tighter bound on the negative entropy production rate.
The difference between the two upper bounds reads
\begin{align}
D_S\dfrac{1-\eta^X_S}{\eta^X_S}-D_I\eta^X_S(1-\eta^X_S)=\dfrac{1-\eta^X_S}{\eta^X_S}D_I\left[\dfrac{D_S}{D_I}-(\eta^X_S)^2\right].
\label{tightness of the bounds}
\end{align}
Therefore, the bound (\ref{trade-off between entropy production and efficiency}) is tighter than (\ref{trade-off between entropy production and efficiency via D_I}) when $\sqrt{D_S/D_I}<\eta^X_S\le1$, while (\ref{trade-off between entropy production and efficiency via D_I}) becomes tighter than (\ref{trade-off between entropy production and efficiency}) when $0\le\eta^X_S<\sqrt{D_S/D_I}$.
Note that $D_S/D_I$ may depend on $\eta^X_S$.

In the linear response regime with $\dot{S}^X_{\mathrm{env}}\le0$ and $\dot{I}^X\le0$, we can prove the input-output fluctuation inequality $D_S\le D_I$ (for the derivation, see Sec.~\ref{Input-output fluctuation inequalities}).
Beyond the linear response regime, however, the input-output fluctuation inequality can be violated, i.e., $D_S$ can become larger than $D_I$~\cite{shiraishi2022there}.

\subsubsection{Trade-off between power and efficiency\label{Trade-off between power and efficiency}}
While we have focused on the negative entropy production rate to characterize the performance of an information-thermodynamic engine, we can also derive a trade-off relation in terms of power, i.e., output work produced per unit of time. 
To define power, we assume that the transition rates satisfy the local detailed balance condition of the following form~\cite{peliti2021stochastic}:
\begin{align}
\ln\dfrac{w^y_{xx'}}{w^y_{x'x}}=\beta\left(\epsilon_{x'y}-\epsilon_{xy}+\Delta^y_{xx'}\right),
\end{align}
where $\beta=(k_{\mathrm{B}}T)^{-1}$ denotes the inverse temperature, $\epsilon_{xy}$ denotes the energy of the state $(x,y)$, and $\Delta^y_{xx'}$ denotes the energy provided by an external agent during the transition $(x',y)\rightarrow(x,y)$.
Then, the average rate of heat absorbed by $X$ from the environment is identified as
\begin{align}
\dot{Q}^X=-k_{\mathrm{B}}T\sum_x\sum_{x'(\neq x)}\sum_yw^y_{xx'}p_t(x',y)\ln\dfrac{w^y_{xx'}}{w^y_{x'x}}.
\end{align}
Similarly, the average rate of work done by the external agent to $X$ is identified as
\begin{align}
\dot{W}^X:=\sum_x\sum_{x'(\neq x)}\sum_yw^y_{xx'}p_t(x',y)\Delta^y_{xx'}.
\end{align}
Finally, the average rate of change of internal energy reads
\begin{align}
\dot{E}^X:=\sum_x\sum_{x'(\neq x)}\sum_yw^y_{xx'}p_t(x',y)\left(\epsilon_{xy}-\epsilon_{x'y}\right).
\end{align}
If we regard $x$ as an externally manipulated control parameter driving $Y$, then $\dot{E}^X$ can also be identified as the power delivered from $X$ to $Y$~\cite{ehrich2023energy,leighton2023inferring}:
\begin{align}
\dot{E}^X=\dot{W}^{X\rightarrow Y}.
\end{align}
Similarly, we can define $\dot{W}^Y$, $\dot{Q}^Y$, and $\dot{W}^{Y\rightarrow X}$.
Then, the first law of stochastic thermodynamics for each subsystem can be expressed as follows (in an averaged form):
\begin{align}
\dot{W}^{X\rightarrow Y}&=\dot{W}^X+\dot{Q}^X,\\
\dot{W}^{Y\rightarrow X}&=\dot{W}^Y+\dot{Q}^Y.
\end{align}
By using these relations, we can rewrite the second law of information thermodynamics in the steady state as
\begin{align}
\beta\dot{W}^X-\beta\dot{W}^{X\rightarrow Y}-\dot{I}^X&\ge0,\\
% \beta\dot{W}^X+\beta\dot{W}^{Y\rightarrow X}+\dot{I}^Y&\ge0,\\
\beta\dot{W}^Y-\beta\dot{W}^{Y\rightarrow X}-\dot{I}^Y&\ge0.
\end{align}
Here, we have assumed that both $X$ and $Y$ are each in contact with a thermal bath at temperature $T$, while the extension to the case of different temperatures is straightforward~\cite{grelier2023unlocking}. 
Note that $\dot{W}^{X\rightarrow Y}=-\dot{W}^{Y\rightarrow X}$ and $\dot{I}^X=-\dot{I}^Y$ in the steady state.
Therefore, $\dot{W}^X$ and $\dot{W}^Y$ cannot both be negative.

Now, suppose that $X$ operates as an information-thermodynamic engine, i.e., $\dot{W}^Y>0$ and $\dot{W}^X<0$.
In this case, we can introduce the following efficiency:
\begin{align}
\eta^X_W:=\dfrac{|\beta\dot{W}^X|}{\beta\dot{W}^{Y\rightarrow X}+\dot{I}^Y},
\end{align}
which satisfies $0\le\eta^X_W\le1$, as can be seen from the second law of information thermodynamics.
The denominator $\beta\dot{W}^{Y\rightarrow X}+\dot{I}^Y=-\beta\dot{W}^{X\rightarrow Y}-\dot{I}^X\ge0$ is called the \textit{transduced capacity}~\cite{lathouwers2022internal,ehrich2023energy}, because it constraints the conversion of the input power $\dot{W}^Y$ into the output power $|\dot{W}^X|$ as $\beta\dot{W}^Y\ge\beta\dot{W}^{Y\rightarrow X}+\dot{I}^Y\ge|\beta\dot{W}^X|$.
The efficiency $\eta^X_W$ quantifies how efficiently $X$ converts the transduced capacity into the output power $|\dot{W}^X|$.

Now we derive a trade-off relation between the output power and the efficiency $\eta^X_W$ by using the bipartite TUR (\ref{Trade-off TUR}).
Let us choose the stochastic work as current $\hat{\mathcal{J}}_\mathcal{T}$ in (\ref{Trade-off TUR}):
\begin{align}
\hat{\mathcal{J}}_\mathcal{T}&=\Delta\hat{W}^X\notag\\
&:=\sum_{x}\sum_{x'(\neq x)}\sum_y\hat{n}^y_{xx'}\Delta^y_{xx'}.
\end{align}
Then, the bipartite TUR gives
\begin{align}
|\dot{W}^X|\le\dfrac{D_W}{k_{\mathrm{B}}T}\dfrac{1-\eta^X_W}{\eta^X_W},
\label{trade off power information-thermodynamic efficiency}
\end{align}
where $D_W$ denotes the fluctuation of the output work defined by
\begin{align}
D_W:=\lim_{\mathcal{T}\rightarrow\infty}\dfrac{1}{2\mathcal{T}}\mathrm{Var}[\Delta\hat{W}^X].
\end{align}
The inequality (\ref{trade off power information-thermodynamic efficiency}) states that an information engine with a finite output power cannot achieve $\eta^X_W=1$ as long as the fluctuation $D_W$ is finite.
\\

\subsection{Memory: $0< \dot{I}^X \le \dot{S}^X_{\mathrm{env}}$   \label{In the case of positive information flow}}
Although we have assumed that $X$ evolves much slower than $Y$, there may be a situation where $X$ measures the state of $Y$, i.e., $\dot{I}^X>0$. 
Even in this case, we can also prove similar trade-off relations concerning the performance of the memory $X$.
We first note that both the entropy production and the information flow associated with $X$ are positive and satisfy the relation $0< \dot{I}^X \le \dot{S}^X_{\mathrm{env}}$.
Then, we can introduce the following information-thermodynamic efficiency:
\begin{align}
\eta^X_I:=\dfrac{\dot{I}^X}{\dot{S}^X_{\mathrm{env}}},
\label{efficiency_2}
\end{align}
which satisfies $0\le\eta^X_I\le1$.
In contrast to $\eta^X_S$, this efficiency quantifies how efficiently $X$ gains information about $Y$ relative to the energy dissipation or thermodynamic cost.
Now we choose the time-integrated stochastic information flow as current $\hat{\mathcal{J}}_\mathcal{T}$ in the bipartite TUR (\ref{Trade-off TUR}).
By noting the positivity of $\dot{S}^X_{\mathrm{env}}$ and $\dot{I}^X$, the bipartite TUR gives the following inequality:
\begin{align}
\dot{I}^X\le D_I\dfrac{1-\eta^X_I}{\eta^X_I}.
\end{align}
This inequality states that a memory with a finite information flow can never attain $\eta^X_I=1$ as long as $D_I$ is finite.

If we choose the stochastic entropy production as current, $\hat{\mathcal{J}}_\mathcal{T}=\Delta\hat{S}^X_{\mathrm{env}}$, then we can obtain a similar trade-off relation where the information flow is bounded by the fluctuation of the stochastic entropy production $D_S$ instead of $D_I$:
\begin{align}
\dot{I}^X\le D_S\eta^X_I(1-\eta^X_I).
\end{align}

\section{Gallavotti-Cohen symmetry and input-output fluctuation inequalities\label{Gallavotti-Cohen symmetry and input-output fluctuation inequalities}}
In this section, we prove that the Gallavotti-Cohen symmetry~\cite{gallavotti1995dynamical,kurchan1998fluctuation,lebowitz1999gallavotti} is satisfied in the fast relaxation limit of $Y$.
As a consequence of this symmetry, we can further show that the input-output fluctuation inequalities hold in the linear response regime even in the presence of an information flow.
\vspace{5mm}

\subsection{Gallavotti-Cohen symmetry\label{Gallavotti-Cohen symmetry}}
Let $\mu(\lambda_S,\lambda_I)$ be the scaled cumulant generating function of the time-integrated currents $\Delta \hat{S}^X_{\mathrm{env}}$ and $\Delta \hat{I}^X$ defined by
\begin{align}
\mu(\lambda_S,\lambda_I):=\lim_{\mathcal{T}\rightarrow\infty}\dfrac{1}{\mathcal{T}}\ln\left\langle e^{\lambda_S\Delta \hat{S}^X_{\mathrm{env}}-\lambda_I\Delta \hat{I}^X}\right\rangle,
\end{align}
where $\lambda_S$ and $\lambda_I$ are the counting fields for $\Delta \hat{S}^X_{\mathrm{env}}$ and $\Delta \hat{I}^X$, respectively.
In this section, we prove that $\mu(\lambda_S,\lambda_I)$ satisfies the following Gallavotti-Cohen symmetry in the fast relaxation limit of $Y$:
\begin{align}
\mu(\lambda_S,\lambda_I)=\mu(-\lambda_S-1,-\lambda_I-1).
% \label{GC symmetry}
\end{align}

To prove this, we first note that $\mu(\lambda_S,\lambda_I)$ can be rewritten as
\begin{align}
\mu(\lambda_S,\lambda_I)=\lim_{\mathcal{T}\rightarrow\infty}\dfrac{1}{\mathcal{T}}\ln\sum_x\sum_yG_{\mathcal{T}}(x,y),
\end{align}
where $G_{\mathcal{T}}(x,y)$ denotes the generating function conditioned to a final state $(x,y)$:
\begin{align}
G_{\mathcal{T}}(x,y)&:=\int d\Delta S^X_{\mathrm{env}}d\Delta I^Xp_\mathcal{T}(x,y,\Delta S^X_{\mathrm{env}},\Delta I^X)\notag\\
&\quad\times e^{\lambda_S\Delta S^X_{\mathrm{env}}-\lambda_I\Delta I^X},
\end{align}
where $p_\mathcal{T}(x,y,\Delta S^X_{\mathrm{env}},\Delta I^X)$ denotes the joint probability density such that the state of the system at time $\mathcal{T}$ is $(x,y)$ and the entropy production and information flow generated up to that time are $\Delta S^X_{\mathrm{env}}$ and $\Delta I^X$, respectively.
Therefore, the property of the scaled cumulant generating function $\mu(\lambda_S,\lambda_I)$ is encoded in the property of the time evolution equation of the generating function $G_{\mathcal{T}}(x,y)$.
The time evolution equation of $G_{\mathcal{T}}(x,y)$ can be obtained by noting that the time evolution equation of $p_\mathcal{T}(x,y,\Delta S^X_{\mathrm{env}},\Delta I^X)$ reads
\begin{widetext}
\begin{align}
\partial_\tau p_\tau(x,y,\Delta S^X_{\mathrm{env}},\Delta I^X)&=\sum_{x'(\neq x)}\left[\widetilde{w}^y_{xx'}p_\tau\left(x',y,\Delta S^X_{\mathrm{env}}-\ln\dfrac{\widetilde{w}^y_{xx'}}{\widetilde{w}^y_{x'x}},\Delta I^X-\ln\dfrac{\pi_{\mathrm{ss}}(y|x)}{\pi_{\mathrm{ss}}(y|x')}\right)-\widetilde{w}^y_{x'x}p_\tau(x,y,\Delta S^X_{\mathrm{env}},\Delta I^X)\right]\notag\\
&\quad+\dfrac{1}{\epsilon}\sum_{y'(\neq y)}\left[\widetilde{w}^{yy'}_xp_\tau(x,y',\Delta S^X_{\mathrm{env}},\Delta I^X)-\widetilde{w}^{y'y}_xp_\tau(x,y,\Delta S^X_{\mathrm{env}},\Delta I^X)\right],
\end{align}
where we have used the dimensionless slow time $\tau:=\mathcal{T}/\tau_X$ and dimensionless transition rates $\widetilde{w}^y_{xx'}:=\tau_Xw^y_{xx'}$ and $\widetilde{w}^{yy'}_x:=\tau_Yw^{yy'}_x$.
Then, we find that the time evolution of $G_\tau(x,y)$ is described by the following tilted dynamics:
\begin{align}
\partial_{\tau}G_{\tau}(x,y)&=\sum_{x'}\left[\mathcal{L}^X_{\lambda_S,\lambda_I}\right]^y_{xx'}G_{\tau}(x',y)+\dfrac{1}{\epsilon}\sum_{y'}\left[\mathcal{L}^Y_{\lambda_S,\lambda_I}\right]^{yy'}_xG_{\tau}(x,y'),
% \partial_{\tau}G_{\tau}(x,y)&=\sum_{x'(\neq x)}\left[\widetilde{w}^y_{xx'}\exp\left(\lambda_S\ln\dfrac{\widetilde{w}^y_{xx'}}{\widetilde{w}^y_{x'x}}-\lambda_I\ln\dfrac{\pi_{\mathrm{ss}}(y|x)}{\pi_{\mathrm{ss}}(y|x')}\right)G_{\tau}(x',y)-\widetilde{w}^y_{x'x}G_{\tau}(x,y)\right]\notag\\
% &\qquad+\dfrac{1}{\epsilon}\sum_{y'(\neq y)}\left[\widetilde{w}^{yy'}_xG_{\tau}(x,y')-\widetilde{w}^{y'y}_xG_{\tau}(x,y)\right].
\label{GC:time evolution of G}
\end{align}
where $\mathcal{L}^X_{\lambda_S,\lambda_I}$ and $\mathcal{L}^Y_{\lambda_S,\lambda_I}$ denote the tilted generators given by
\begin{align}
\left[\mathcal{L}^X_{\lambda_S,\lambda_I}\right]^y_{xx'}:=
\begin{cases}
\widetilde{w}^y_{xx'}\exp\left(\lambda_S\ln\dfrac{\widetilde{w}^y_{xx'}}{\widetilde{w}^y_{x'x}}-\lambda_I\ln\dfrac{\pi_{\mathrm{ss}}(y|x)}{\pi_{\mathrm{ss}}(y|x')}\right)\quad&(x\neq x'),\\
-\sum_{x'(\neq x)}\widetilde{w}^y_{x'x}\quad&(x=x'),
\end{cases}
\end{align}
\begin{align}
\left[\mathcal{L}^Y_{\lambda_S,\lambda_I}\right]^{yy'}_x:=
\begin{cases}
\widetilde{w}^{yy'}_x\quad&(y\neq y'),\\
-\sum_{y'(\neq y)}\widetilde{w}^{y'y}_x\quad&(y=y').
\end{cases}
\end{align}
\end{widetext}
We now assume that $G_\tau$ has asymptotic expansions in terms of the asymptotic sequences $\{\epsilon^n\}^\infty_{n=0}$ as $\epsilon\rightarrow0$:
\begin{align}
G_\tau=G^{(0)}_\tau+\epsilon G^{(1)}_\tau+\cdots.
\end{align}
Here, we impose the normalization condition
\begin{align}
\sum_yG^{(0)}_\tau(x,y)&=\sum_yG_\tau(x,y)\notag\\
&=:G^X_\tau(x).
\end{align}
By substituting this expansion into (\ref{GC:time evolution of G}), we find that the leading order gives
\begin{align}
\sum_{y'}\left[\mathcal{L}^Y_{\lambda_S,\lambda_I}\right]^{yy'}_xG^{(0)}_{\tau}(x,y')=0.
% \sum_{y'(\neq y)}\left[\widetilde{w}^{yy'}_xG^{(0)}_{\tau}(x,y')-\widetilde{w}^{y'y}_xG^{(0)}_{\tau}(x,y)\right]=0.
\end{align}
From the Perron-Frobenius theorem and the normalization condition, we find that $G^{(0)}_{\tau}$ has the form
\begin{align}
G^{(0)}_{\tau}(x,y)=G^X_{\tau}(x)\pi_{\mathrm{ss}}(y|x).
\end{align}
The subleading order of (\ref{GC:time evolution of G}) yields
\begin{align}
\partial_{\tau}G^{(0)}_{\tau}(x,y)&=\sum_{x'}\left[\mathcal{L}^X_{\lambda_S,\lambda_I}\right]^y_{xx'}G^{(0)}_{\tau}(x',y)\notag\\
&\quad+\sum_{y'}\left[\mathcal{L}^Y_{\lambda_S,\lambda_I}\right]^{yy'}_xG^{(1)}_{\tau}(x,y').
% \partial_{\tau}G^{(0)}_{\tau}(x,y)&=\sum_{x'(\neq x)}\left[\widetilde{w}^y_{xx'}\exp\left(\lambda_S\ln\dfrac{\widetilde{w}^y_{xx'}}{\widetilde{w}^y_{x'x}}-\lambda_I\ln\dfrac{\pi_{\mathrm{ss}}(y|x)}{\pi_{\mathrm{ss}}(y|x')}\right)G^{(0)}_{\tau}(x',y)-\widetilde{w}^y_{x'x}G^{(0)}_{\tau}(x,y)\right]\notag\\
% &\qquad+\sum_{y'(\neq y)}\left[\widetilde{w}^{yy'}_xG^{(1)}_{\tau}(x,y')-\widetilde{w}^{y'y}_xG^{(1)}_{\tau}(x,y)\right].
% \label{GC:time evolution of G}
\end{align}
From the solvability condition for $G^{(1)}_\tau$, we obtain the effective dynamics for $G^X_\tau(x)$:
\begin{align}
\partial_\tau G^X_{\tau}(x)=\sum_{x'}\left[\overline{\mathcal{L}}^X_{\lambda_S,\lambda_I}\right]_{xx'}G^X_\tau(x'),
% \partial_\tau G^X_{\tau}(x)=\sum_{x'(\neq x)}\left\{\sum_y\left[\widetilde{w}^y_{xx'}\exp\left(\lambda_S\ln\dfrac{\widetilde{w}^y_{xx'}}{\widetilde{w}^y_{x'x}}-\lambda_I\ln\dfrac{\pi_{\mathrm{ss}}(y|x)}{\pi_{\mathrm{ss}}(y|x')}\right)\pi_{\mathrm{ss}}(y|x')\right]G^X_\tau(x')-\sum_y\left[\widetilde{w}^y_{x'x}\pi_{\mathrm{ss}}(y|x)\right]G^X_\tau(x)\right\}.
\end{align}
where $\overline{\mathcal{L}}^X_{\lambda_S,\lambda_I}$ denotes the effective tilted generator given by
\begin{widetext}
\begin{align}
\left[\overline{\mathcal{L}}^X_{\lambda_S,\lambda_I}\right]_{xx'}&:=\sum_y\left[\mathcal{L}^X_{\lambda_S,\lambda_I}\right]^y_{xx'}\pi_{\mathrm{ss}}(y|x')\notag\\
&=
\begin{cases}
\sum_y\left[\widetilde{w}^y_{xx'}\exp\left(\lambda_S\ln\dfrac{\widetilde{w}^y_{xx'}}{\widetilde{w}^y_{x'x}}-\lambda_I\ln\dfrac{\pi_{\mathrm{ss}}(y|x)}{\pi_{\mathrm{ss}}(y|x')}\right)\pi_{\mathrm{ss}}(y|x')\right]\quad&(x\neq x'),\\
-\sum_{x'(\neq x)}\left[\sum_y\widetilde{w}^y_{x'x}\pi_{\mathrm{ss}}(y|x)\right]\quad&(x=x').
\end{cases}
\end{align}
\end{widetext}

Importantly, this effective tilted generator satisfies the following property:
\begin{align}
(\overline{\mathcal{L}}^X_{\lambda_S,\lambda_I})^\top=\overline{\mathcal{L}}^X_{-\lambda_S-1,-\lambda_I-1},
\end{align}
where $\top$ denotes the matrix transpose.
Because the scaled cumulant generating function is equal to the largest eigenvalue of this effective tilted generator, the Gallavotti-Cohen symmetry follows from this property:
\begin{align}
\mu(\lambda_S,\lambda_I)=\mu(-\lambda_S-1,-\lambda_I-1).
\label{GC symmetry}
\end{align}

\subsection{Input-output fluctuation inequalities\label{Input-output fluctuation inequalities}}
In the linear response regime, where the scaled cumulant generating function can be approximated by a quadratic form~\cite{peliti2021stochastic}, the Gallavotti-Cohen symmetry (\ref{GC symmetry}) constrains its form as
\begin{align}
\mu(\lambda_S,\lambda_I)&=a\left(\lambda_S+\dfrac{1}{2}\right)^2+b\left(\lambda_S+\dfrac{1}{2}\right)\left(\lambda_I+\dfrac{1}{2}\right)\notag\\
&\quad+c\left(\lambda_I+\dfrac{1}{2}\right)^2-\dfrac{1}{4}(a+b+c),
\end{align}
where $a, b, c$ are constants.
From the convexity of $\mu(\lambda_S,\lambda_I)$, these coefficients satisfy $a\ge0$, $c\ge0$, and $ac-b^2/4\ge0$.
By noting that
\begin{align}
\dot{S}^X_{\mathrm{env}}&=\left.\dfrac{\partial}{\partial \lambda_S}\mu(\lambda_S,\lambda_I)\right|_{\lambda_S,\lambda_I=0}=a+\dfrac{b}{2},\\
\dot{I}^X&=-\left.\dfrac{\partial}{\partial \lambda_I}\mu(\lambda_S,\lambda_I)\right|_{\lambda_S,\lambda_I=0}=-c-\dfrac{b}{2},
\end{align}
these coefficients are further constrained by the second law of information thermodynamics to satisfy $a+b+c\ge0$.

\subsubsection{Information-thermodynamic engine: $0< -  \dot{S}^X_{\mathrm{env}} \le - \dot{I}^X$}
We consider the case of $-\dot{I}^X\ge\dot{S}^X_{\mathrm{env}}$, i.e., $c\ge a$, which includes the case where $X$ acts as an information-thermodynamic engine with $0<-\dot{S}^X_{\mathrm{env}}\le-\dot{I}^X$. Since we have
\begin{align}
D_S&=\dfrac{1}{2}\left.\dfrac{\partial^2}{\partial \lambda^2_S}\mu(\lambda_S,\lambda_I)\right|_{\lambda_S,\lambda_I=0}=a,\\
D_I&=\dfrac{1}{2}\left.\dfrac{\partial^2}{\partial \lambda^2_I}\mu(\lambda_S,\lambda_I)\right|_{\lambda_S,\lambda_I=0}=c,
\end{align}
these relations between the coefficients $a, b, c$ lead to the following input-output fluctuation inequalities:
\begin{align}
D_S&\le D_I,\label{Input-output inequality_1}\\
\dfrac{D_I}{(\dot{I}^X)^2}&\le \dfrac{D_S}{(\dot{S}^X_{\mathrm{env}})^2}.\label{Input-output inequality_2}
\end{align}
These inequalities state that the fluctuation of the output current (negative entropy production) is smaller than that of the input current (information flow), while the relative fluctuation of the output current is larger than that of the input current.

\subsubsection{Memory: $0< \dot{I}^X \le \dot{S}^X_{\mathrm{env}}$}
We can also derive input-output fluctuation inequalities when $-\dot{I}^X\le\dot{S}^X_{\mathrm{env}}$, i.e., $c\le a$, which includes the case where $X$ plays the role of a memory with $0< \dot{I}^X \le \dot{S}^X_{\mathrm{env}}$.
In this case, the information flow $\dot{I}^X$ corresponds to the output current while the entropy production rate $\dot{S}^X_{\mathrm{env}}$ corresponds to the input current.
Obviously, we have the following relations:
\begin{align}
D_S&\ge D_I,\label{memo1}\\
\dfrac{D_I}{(\dot{I}^X)^2}&\ge \dfrac{D_S}{(\dot{S}^X_{\mathrm{env}})^2}.\label{memo2}
\end{align}

\section{Examples\label{Examples}}
In this section, we illustrate our results, the trade-offs for information-thermodynamic engines and the input-output fluctuation inequalities, using two simple examples.
The first example is coupled quantum dots, which is one of the simplest models of autonomous Maxwell's demon~\cite{strasberg2013thermodynamics,horowitz2014thermodynamics}.
As a second example, we consider coupled linear overdamped Langevin equations, which ubiquitously appear in biological contexts with the linear noise approximation~\cite{tuanase2006signal,thomas2012slow,maity2015role,ito2015maxwell}. 
Interestingly, the equality condition of the trade-offs (\ref{trade-off between entropy production and efficiency}) and (\ref{trade-off between entropy production and efficiency via D_I}) is satisfied even far from equilibrium in this case.

\subsection{Coupled quantum dots\label{Coupled quantum-dots}}
%%%%%% Coupled quantum-dots
\subsubsection{Model}
We consider the system composed of two single-level quantum dots $X$ and $Y$.
Let $x\in\{0,1\}$ and $y\in\{0,1\}$ be occupation variables on each particle site, where $x=1$ and $y=1$ ($x=0$ and $y=0$) represent that the site of $X$ and $Y$ is filled (empty), respectively.
The energy of $X$ is $\epsilon_X$ when it is filled with a particle and zero when it is empty.
A single particle site of $X$ exchanges particles with two particle reservoirs $\nu=L, R$ at temperature $T$ and chemical potential $\mu_\nu$.
We assume that $\Delta\mu:=\mu_L-\mu_R>0$.
Let $p_t(x,y)$ be the probability of state $(x,y)$ at time $t$.
The time evolution of $p_t(x,y)$ is described by the master equation:
\begin{multline}
\partial_tp_t(x,y)=\sum_{\nu}\left[w^{(\nu)y}_{xx'}p_t(x',y)-w^{(\nu)y}_{x'x}p_t(x,y)\right]\\
+w^{yy'}_xp_t(x,y')-w^{y'y}_xp_t(x,y),
\label{quantum dot}
\end{multline}
where $x':=1-x$ and $y':=1-y$.
Here, $w^{(\nu)y}_{xx'}$ denotes the time-independent transition rate from $x'$ to $x$ induced by the reservoir $\nu$, which satisfies the local detailed balance condition:
\begin{align}
\dfrac{w^{(\nu)y}_{10}}{w^{(\nu)y}_{01}}=\exp(-\beta(\epsilon_X-\mu_\nu)).
\end{align}
We suppose that the transition rates have the form
\begin{align}
w^{(L)0}_{10}&=\tilde{\Gamma}_Xf_L,\quad&w^{(R)0}_{10}&=\Gamma_X f_R,\\
w^{(L)1}_{10}&=\Gamma_X f_L,\quad&w^{(R)1}_{10}&=\tilde{\Gamma}_Xf_R,\\
w^{(L)0}_{01}&=\tilde{\Gamma}_X(1-f_L),\quad&w^{(R)0}_{01}&=\Gamma_X(1-f_R),\\
w^{(L)1}_{01}&=\Gamma_X(1-f_L),\quad&w^{(R)1}_{01}&=\tilde{\Gamma}_X(1-f_R),
\end{align}
where $f_\nu:=[\exp(\beta(\epsilon_X-\mu_\nu))+1]^{-1}$ is the Fermi distribution function, and $\Gamma_X$ ($\tilde{\Gamma}_X$) denotes a positive coupling strength.
Below, we focus on the case where $\tilde{\Gamma}_X\ll\Gamma_X$.
The above form of transition rates implies that the coupling strength of the $R$($L$)-reservoir changes from $\Gamma_X$ to $\tilde{\Gamma}_X$ when $Y$ is filled (empty) with a particle.
The transition rates associated with $Y$ are given as follows:
\begin{align}
w^{yy'}_x=
\begin{cases}
\Gamma_Y\varepsilon\qquad &(y,y')=(1-x,x),\\
\Gamma_Y(1-\varepsilon)\qquad &(y,y')=(x,1-x),
\end{cases}
\end{align}
where $\Gamma_Y$ is a coupling strength, and $\varepsilon$ can be interpreted as an error probability with $0\le\varepsilon\le1$.

\begin{figure}[t]
\includegraphics[width=8.6cm]{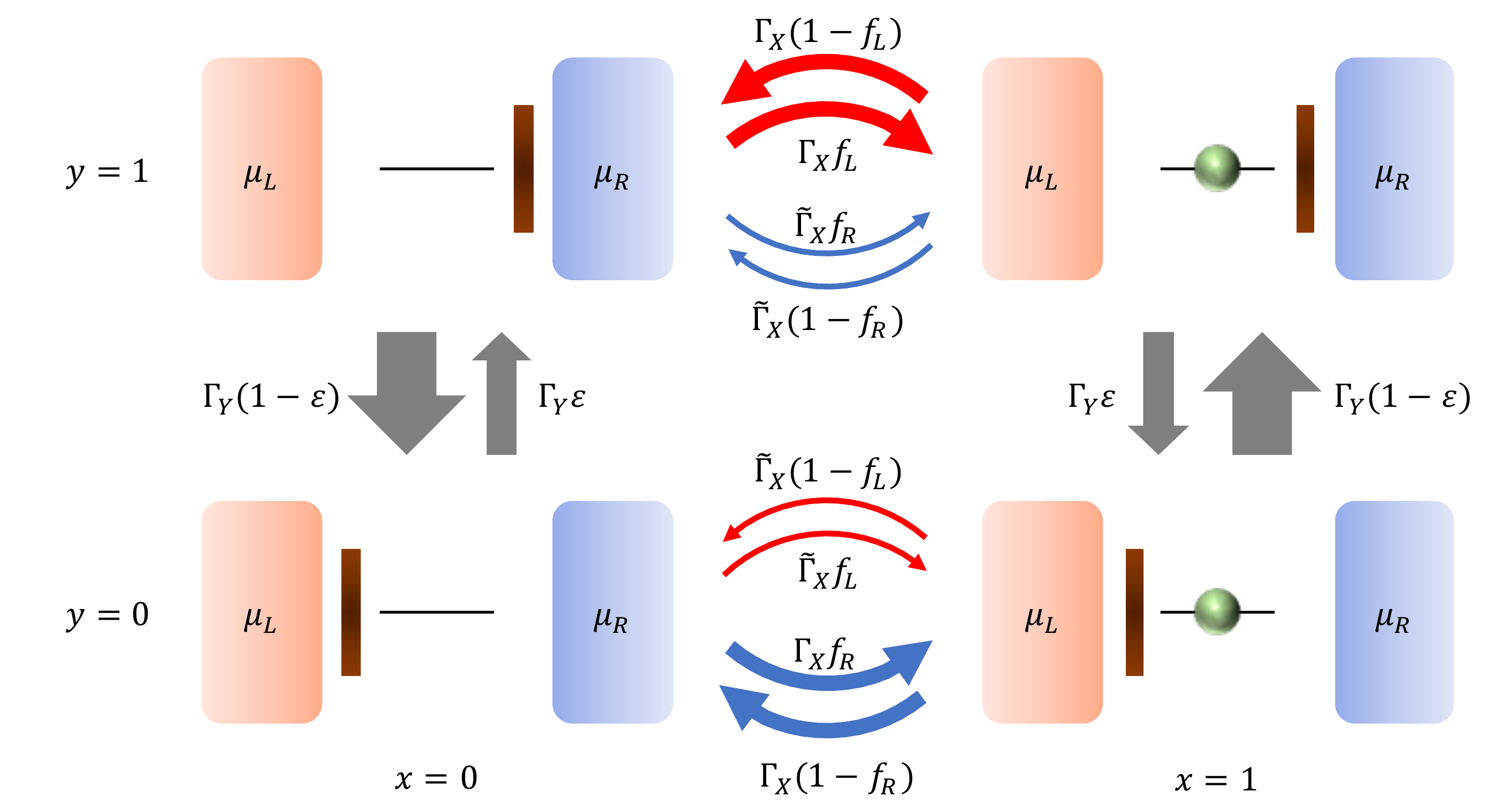}
\caption{Schematic of the coupled quantum dots.
The single particle site of $X$ exchanges particles (green dot) with the two particle reservoirs at a chemical potential $\mu_L$ and $\mu_R$.
The position of the wall represents the state of $Y$: the wall is inserted on the left side when $y=0$, while it is inserted on the right side when $y=1$.
The red and blue arrows represent the transition rates $w^{(L)y}_{xx'}$ and $w^{(R)y}_{xx'}$, respectively.
The gray arrows represent the transition rates associated with $Y$, $w^{yy'}_x$.
The thickness of these arrows indicates the magnitude of each transition rate.}
\label{fig:coupled_quantum_dot}
\end{figure}

In this model, the subsystem $Y$ acts as Maxwell's demon when $\varepsilon$ is sufficiently small.
To understand this point intuitively, let us consider the state of $Y$ as representing the position of the wall, which is inserted between the single site of $X$ and the reservoir.
In other words, when $y=0$ ($y=1$), the wall is inserted between the site of $X$ and the $L$ ($R$) reservoir and prohibits the transition due to the $L$ ($R$) reservoir by changing the coupling strength from $\Gamma_X$ to $\tilde{\Gamma}_X$ (see Fig.~\ref{fig:coupled_quantum_dot}).
As a result, particles are transferred from the $R$ to $L$ reservoirs against the chemical potential difference.

\subsubsection{Fast relaxation limit of $Y$}
Hereafter, we focus on the case where $Y$ is faster than $X$, i.e., $\Gamma_Y\gg\Gamma_X\gg\tilde{\Gamma}_X$.
By performing a perturbation expansion following Sec.~\ref{Fast relaxation limit for Y}, we can show that $p_t(x,y)\simeq p^X_t(x)\pi_{\mathrm{ss}}(y|x)$ with
\begin{align}
\pi_{\mathrm{ss}}(y=x|x)&=1-\varepsilon,\\
\pi_{\mathrm{ss}}(y=1-x|x)&=\varepsilon.
\end{align}
The effective dynamics for $X$ is then given by
\begin{align}
\partial_t p^X_t(x)\simeq\sum_{\nu}\left[\overline{w}^{(\nu)}_{xx'}p^X_t(x')-\overline{w}^{(\nu)}_{x'x}p^X_t(x)\right],
\end{align}
where $\overline{w}^{(\nu)}_{xx'}:=\sum_yw^{(\nu)y}_{xx'}\pi_{\mathrm{ss}}(y|x')$ denotes the effective transition rates:
\begin{align}
\overline{w}^{(L)}_{10}&=\left[\varepsilon\Gamma_X+(1-\varepsilon)\tilde{\Gamma}_X\right]f_L,\\
\overline{w}^{(L)}_{01}&=\left[(1-\varepsilon)\Gamma_X+\varepsilon\tilde{\Gamma}_X\right](1-f_L),\\
\overline{w}^{(R)}_{10}&=\left[(1-\varepsilon)\Gamma_X+\varepsilon\tilde{\Gamma}_X\right]f_R,\\
\overline{w}^{(R)}_{01}&=\left[\varepsilon\Gamma_X+(1-\varepsilon)\tilde{\Gamma}_X\right](1-f_R).
\end{align}

Thus, in the fast relaxation limit of $Y$, the system $X$ can be considered as an autonomous system where the coupling strength of the reservoirs changes autonomously.
More specifically, when $x=1$, the coupling strength of the $R$-reservoir changes from the original strength, $\Gamma_X$, to a smaller value, $\tilde{\Gamma}_X$, while that of the $L$-reservoir remains unchanged.
In contrast, when $x=0$, the coupling strength of the $L$-reservoir becomes small while that of the $R$-reservoir remains at the original strength $\Gamma_X$.
This autonomous control is probabilistic and has the error probability $\varepsilon$.

\subsubsection{Trade-off between power and efficiency}
We first consider the trade-off between the negative entropy production rate and information-thermodynamic efficiency (\ref{trade-off between entropy production and efficiency}).
Note that (\ref{trade-off between entropy production and efficiency}) corresponds to the trade-off between power and efficiency (\ref{trade off power information-thermodynamic efficiency}), because $\dot{S}^X_{\mathrm{env}}=\beta\dot{W}^X$ in this model.

We first calculate the average rate of chemical work $\dot{W}^X$.
By defining $b_{xx'}$ ($x'=1-x$) as
\begin{align}
b_{xx'}:=
\begin{cases}
1\quad&(x=1, x'=0),\\
-1\quad&(x=0, x'=1),
\end{cases}
\end{align}
we note that $b_{xx'}\mu_\nu$ corresponds to the energy provided by the particle reservoir $\nu$ during the transition $(x',y)\rightarrow(x,y)$.
Then, the average rate of chemical work reads
\begin{align}
\dot{W}^X&=\sum_\nu\sum_x\sum_yw^{(\nu)y}_{xx'}p_{\mathrm{ss}}(x',y)b_{xx'}\mu_\nu\notag\\
&\simeq\sum_\nu\sum_x\sum_yw^{(\nu)y}_{xx'}\pi_{\mathrm{ss}}(y|x')p^X_{\mathrm{ss}}(x')b_{xx'}\mu_\nu\notag\\
&=\sum_\nu\sum_x\overline{w}^{(\nu)}_{xx'}p^X_{\mathrm{ss}}(x')b_{xx'}\mu_\nu\notag\\
&=J_X\Delta\mu,
\end{align}
where in the second line, we have used $p_{\mathrm{ss}}(x',y)\simeq\pi_{\mathrm{ss}}(y|x')p^X_{\mathrm{ss}}(x')$ in the fast relaxation limit of $Y$.
In the last line, $J_X$ denotes the net particle current from $L$ to $R$, which is conjugate with the chemical potential difference $\Delta\mu$:
\begin{align}
J_X&=\overline{w}^{(L)}_{10}p^X_{\mathrm{ss}}(0)-\overline{w}^{(L)}_{01}p^X_{\mathrm{ss}}(1)\notag\\
% &=\sum_y\left[w^{(L)y}_{10}\pi_{\mathrm{ss}}(y|0)p^X_{\mathrm{ss}}(0)-w^{(L)y}_{01}\pi_{\mathrm{ss}}(y|1)p^X_{\mathrm{ss}}(1)\right]\notag\\
&=\Gamma_X\dfrac{\varepsilon^2f_L(1-f_R)-(1-\varepsilon)^2(1-f_L)f_R}{1+(2\varepsilon-1)(f_L-f_R)}+O(\tilde{\Gamma}_X).
\label{J_X}
\end{align}
The net particle current $J_X$ becomes negative when $\varepsilon$ is smaller than the critical value $\varepsilon_*$, which can be evaluated as
\begin{align}
\varepsilon_*=\dfrac{(1-f_L)f_R}{f_L-f_R}\left[-1+\sqrt{1+\dfrac{f_L-f_R}{(1-f_L)f_R}}\right]+O\left(\dfrac{\tilde{\Gamma}_X}{\Gamma_X}\right).
\end{align}
Note that $\varepsilon_*<1/2$ because $f_L(1-f_R)>f_R(1-f_L)$, which follows from the condition $\Delta\mu=\mu_L-\mu_R>0$.

Similarly, the information flow can be expressed as
\begin{align}
\dot{I}^X&=\sum_\nu\sum_x\sum_yw^{(\nu)y}_{xx'}p_{\mathrm{ss}}(x',y)\ln\dfrac{p_{\mathrm{ss}}(y|x)}{p_{\mathrm{ss}}(y|x')}\notag\\
&\simeq\sum_\nu\sum_x\sum_yw^{(\nu)y}_{xx'}\pi_{\mathrm{ss}}(y|x')p^X_{\mathrm{ss}}(x')\ln\dfrac{\pi_{\mathrm{ss}}(y|x)}{\pi_{\mathrm{ss}}(y|x')}\notag\\
&=J_IF_I,
\end{align}
where in the second line, we have used $p_{\mathrm{ss}}(x',y)\simeq\pi_{\mathrm{ss}}(y|x')p^X_{\mathrm{ss}}(x')$ in the fast relaxation limit of $Y$.
In the last line, $F_I$ denotes the information affinity defined as
\begin{align}
F_I&:=\ln\dfrac{\pi_{\mathrm{ss}}(0|0)\pi_{\mathrm{ss}}(1|1)}{\pi_{\mathrm{ss}}(0|1)\pi_{\mathrm{ss}}(1|0)}\notag\\
&=2\ln\dfrac{1-\varepsilon}{\varepsilon},
\end{align}
and $J_I$ denotes the probability current that is conjugate with $F_I$:
\begin{align}
J_I&=\sum_\nu\left[w^{(\nu)0}_{01}\pi_{\mathrm{ss}}(0|1)p^X_{\mathrm{ss}}(1)-w^{(\nu)0}_{10}\pi_{\mathrm{ss}}(0|0)p^X_{\mathrm{ss}}(0)\right]\notag\\
&=J_X+O(\tilde{\Gamma}_X).
% &=\sum_\nu\left[w^{(\nu)0}_{01}\pi_{\mathrm{ss}}(0|1)p^X_{\mathrm{ss}}(1)-w^{(\nu)0}_{10}\pi_{\mathrm{ss}}(0|0)p^X_{\mathrm{ss}}(0)\right]\notag\\
% &=\Gamma\dfrac{\epsilon^2 f_L(1-f_R)-(1-\epsilon)^2(1-f_L)f_R}{1+(2\epsilon-1)(f_L-f_R)}+O(\tilde{\Gamma})\notag\\
\end{align}
Thus, the tight-coupling condition is satisfied in the limit $\tilde{\Gamma}_X/\Gamma_X\ll1$.
Since $\varepsilon_*<1/2$, the information flow $\dot{I}^X$ also becomes negative when $\varepsilon<\varepsilon_*$.

The fluctuation of the chemical work can be calculated by considering the tilted dynamics (see Appendix~\ref{Appendix: Coupled quantum dots}).
The result reads
\begin{align}
D_W=D_n\Delta\mu^2,
\label{ex1:D_W}
\end{align}
where $D_n$ denotes the fluctuation of the net particle current:
\begin{align}
D_n&=\dfrac{\Gamma_X}{2}\left\{\dfrac{(1-\varepsilon)^2(1-f_L)f_R+\varepsilon^2f_L(1-f_R)}{1+(2\varepsilon-1)(f_L-f_R)}\right.\notag\\
&\quad\left.-\dfrac{2\left[(1-\varepsilon)^2(1-f_L)f_R-\varepsilon^2f_L(1-f_R)\right]^2}{\left[1+(2\varepsilon-1)(f_L-f_R)\right]^3}\right\}+O(\tilde{\Gamma}_X).
\end{align}
\begin{figure}[t]
\center
\includegraphics[width=8.6cm]{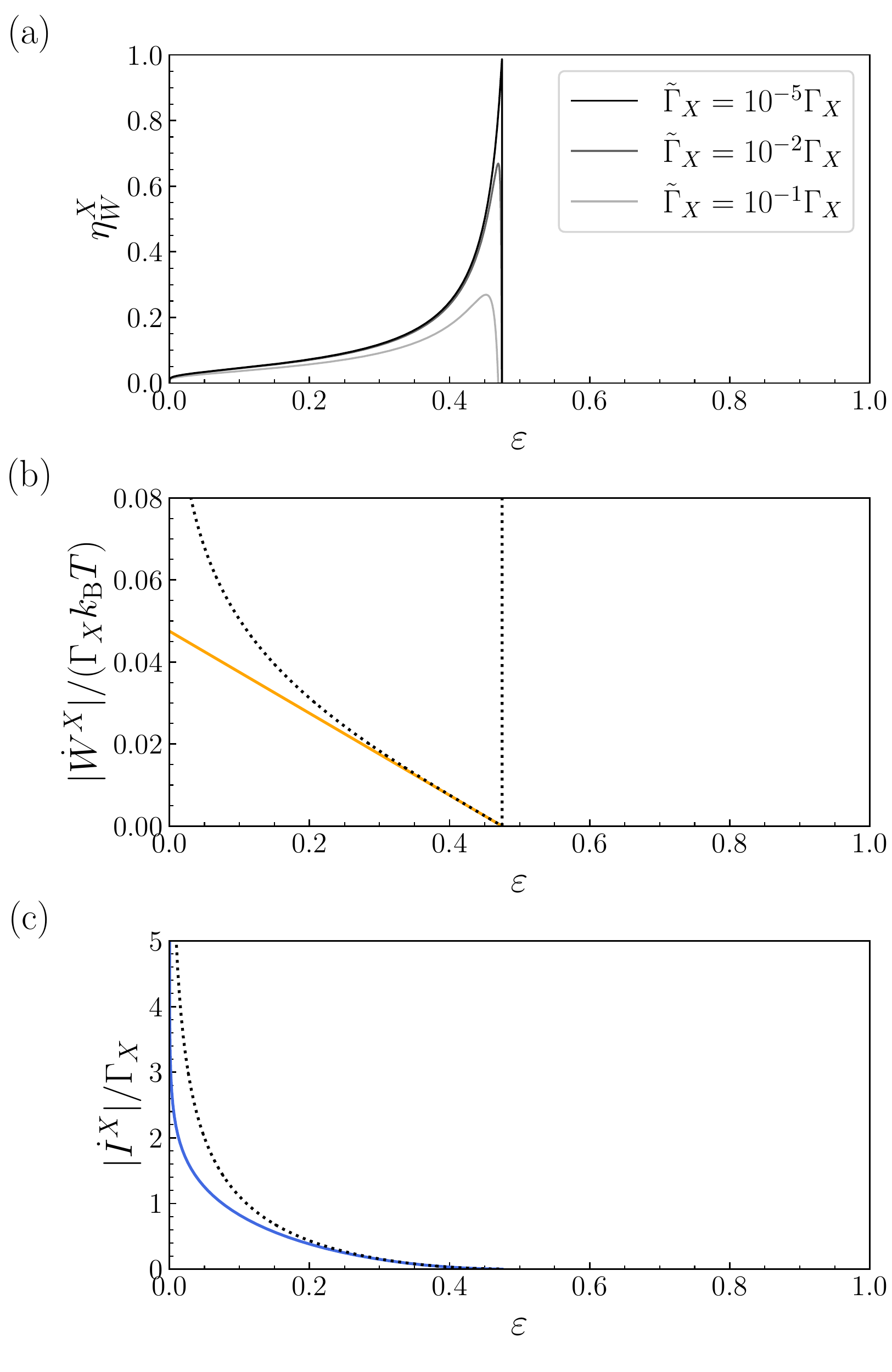}
\caption{(a) $\varepsilon$-dependence of information-thermodynamic efficiency $\eta^X_W$. 
(b) $\varepsilon$-dependence of the power $|\dot{W}^X|$ with $\tilde{\Gamma}_X=10^{-5}\Gamma_X$. The dotted line denotes the upper bound of (\ref{trade off power information-thermodynamic efficiency}). 
(c) $\varepsilon$-dependence of the information flow $|\dot{I}^X|$ with $\tilde{\Gamma}_X=10^{-5}\Gamma_X$. The dotted line denotes the upper bound of (\ref{trade-off between information and efficiency}). In all panels, the parameter values are $\varepsilon_X=1$, $\mu_L=1.1$, $\mu_R=0.9$, and $k_{\mathrm{B}}T=1$.}
\label{fig:ex1_result}
\end{figure}

We now focus on the case of $\varepsilon<\varepsilon_*$, where the system $X$ acts as an information-thermodynamic engine with $\dot{W}^X<0$.
The corresponding information-thermodynamic efficiency reads
\begin{align}
\eta^X_W=\eta^X_S=\dfrac{|J_X|F_X}{|J_I|F_I}\simeq\dfrac{F_X}{F_I}\le1,
\end{align}
where $F_X:=\beta\Delta\mu$ denotes the thermodynamic affinity conjugate with $J_X$.
The $\varepsilon$-dependence of the efficiency $\eta^X_W$ and the output power $|\dot{W}^X|$ is shown in Fig.~\ref{fig:ex1_result}(a) and (b), respectively.
From this figure, we can see that the output power does not remain finite as $\eta^X_W\rightarrow1$.
This result is consistent with the trade-off between power and information-thermodynamic efficiency (\ref{trade off power information-thermodynamic efficiency}) as illustrated in Fig.~\ref{fig:ex1_result}(b).

We next consider the trade-off relation where the negative entropy production is bounded by the fluctuation of the time-integrated stochastic information flow $D_I$ (\ref{trade-off between entropy production and efficiency via D_I}).
In terms of the power $\dot{W}^X$, it can be expressed as
\begin{align}
|\dot{W}^X|\le D_Ik_{\mathrm{B}}T\eta^X_W(1-\eta^X_W).
\label{ex1: trade-off between entropy production and efficiency via D_I}
\end{align}
The fluctuation of the information flow $D_I$ can also be calculated by using the tilted dynamics as
\begin{align}
D_I=D_nF^2_I,
% \label{ex1:D_W}
\end{align}
which satisfies $\beta\sqrt{D_W/D_I}=F_X/F_I\simeq\eta^X_W$.
Therefore, from (\ref{tightness of the bounds}), it follows that the upper bound of (\ref{ex1: trade-off between entropy production and efficiency via D_I}) is exactly the same as that of (\ref{trade off power information-thermodynamic efficiency}) for $\varepsilon<\varepsilon_*$.

For comparison, we also plot the information flow and its upper bound (\ref{trade-off between information and efficiency}) in Fig.~\ref{fig:ex1_result}(c).
As in the case of the output power, the information flow also vanishes as the efficiency $\eta^X_S(=\eta^X_W)$ approaches 1. 
We note that $|\dot{I}^X|\rightarrow\infty$ as $\varepsilon\rightarrow0$ because the information affinity $F_I$ diverges.

\subsubsection{Input-output fluctuation inequalities}
We now consider the input-output fluctuation inequalities for $\varepsilon<\varepsilon_*$, where the entropy production ($\dot{S}^X_{\mathrm{env}}=\beta\dot{W}^X$) and the information flow correspond to the output and input currents, respectively.
Since $F_X/F_I\le1$, we can easily confirm that $D_S\le D_I$ and $D_I/(\dot{I}^X)^2=D_S/(\dot{S}^X_{\mathrm{env}})^2$.
Thus, the input-output fluctuation inequalities are satisfied even beyond the linear response regime in this model.
Furthermore, the equality is achieved for the inequality regarding the relative fluctuations.

\subsection{Coupled linear overdamped Langevin equations\label{Coupled linear overdamped Langevin equations}}
%%%%%% Coupled linear overdamped Langevin
\subsubsection{Model}
We consider the following coupled linear overdamped Langevin equations:
\begin{align}
\dot{x}_t&=-\omega^X x_t+\omega^{XY} y_t+\sqrt{2D^X}\xi^X_t,\label{ex2:model-x}\\
\dot{y}_t&=\omega^{YX} x_t-\omega^Y y_t+\sqrt{2D^Y}\xi^Y_t,\label{ex2:model-y}
\end{align}
where $\xi^Z_t$ ($Z=X,Y$) is a zero-mean white Gaussian noise that satisfies $\langle\xi^Z_t\xi^{Z'}_{t'}\rangle=\delta_{ZZ'}\delta(t-t')$, and $D^Z>0$ denotes the noise intensity.
Here, $x_t$ relaxes exponentially with decay rate $\omega^X>0$ and is affected by $y_t$ with rate $\omega^{XY}$, while $y_t$ also relaxes exponentially with decay rate $\omega^Y>0$ and detects $x_t$ with the differential gain $\omega^{YX}$.
We assume that
\begin{align}
\omega^X\omega^Y-\omega^{XY}\omega^{YX}>0
\label{stationary condition}
\end{align}
to ensure that the system reaches a steady state~\cite{gardiner1985handbook}.
The corresponding Fokker-Planck equation reads
\begin{align}
\partial_tp_t(x,y)&=-\partial_xJ^X_t(x,y)-\partial_yJ^Y_t(x,y),
% &=-\partial_y\left[(-\omega^Y s+\omega^{YX} x)p_t(x,y)-D^Y\partial_yp_t(x,y)\right]\notag\\
% &\qquad-\partial_x\left[(\omega^{XY} s-\omega^X x)p_t(x,y)-D^X\partial_xp_t(x,y)\right].
\end{align}
where $J^X_t(x,y)$ and $J^Y_t(x,y)$ denote the probability currents:
\begin{align}
J^X_t(x,y)&:=(-\omega^X x + \omega^{XY} y)p_t(x,y)-D^X\partial_xp_t(x,y),\\
J^Y_t(x,y)&:=(\omega^{YX} x -\omega^Y y)p_t(x,y)-D^Y\partial_yp_t(x,y).
\end{align}

While this model is exactly solvable, it is widely used to describe biological systems such as signal transduction networks and gene regulatory networks~\cite{tuanase2006signal,thomas2012slow,maity2015role,ito2015maxwell}.
In the context of heat engines, this model includes a Brownian gyrator~\cite{filliger2007brownian} and can be experimentally realized in, e.g., electronic and colloidal systems~\cite{chiang2017electrical,argun2017experimental}.
Note that the system can be far from equilibrium due to the nonreciprocal interactions (when $\omega^{XY}\neq \omega^{YX}$) or the heat flow (when $D^X\neq D^Y$).

\subsubsection{Fast relaxation limit of $Y$}
Hereafter, we focus on the case where $Y$ relaxes much faster than $X$.
We introduce a dimensionless slow time $\tau:=\omega^X t$ and a small parameter $\epsilon:=\omega^X/\omega^Y\ll1$.
Correspondingly, we introduce dimensionless rates $\bar{\omega}^{XY}:=\omega^{XY}/\omega^X$ and $\bar{\omega}^{YX}:=\omega^{YX}/\omega^Y$ and dimensionless noise intensities $\bar{D}^X:=D^X/\omega^X$ and $\bar{D}^Y:=D^Y/\omega^Y$. 
From the condition (\ref{stationary condition}) and the positivity of $\omega^X$ and $\omega^Y$, we note that $\bar{\omega}^{XY}\bar{\omega}^{YX}<1$.
Then, the time evolution equations (\ref{ex2:model-x}) and (\ref{ex2:model-y}) can be rewritten as
\begin{align}
\dot{x}_\tau&=\left[-x_\tau+\bar{\omega}^{XY} y_\tau\right]+\sqrt{2\bar{D}^X}\xi^X_\tau,\label{ex2:model-x_dimensionless}\\
\dot{y}_\tau&=\dfrac{1}{\epsilon}\left[\bar{\omega}^{YX} x_\tau-y_\tau\right]+\sqrt{\dfrac{2\bar{D}^Y}{\epsilon}}\xi^Y_\tau.\label{ex2:model-s_dimensionless}
\end{align}
In the fast relaxation limit $\epsilon\rightarrow0$, the joint probability density $p_\tau(x,y)$ can be approximated as $p_\tau(x,y)\simeq p^X_\tau(x)\pi_{\mathrm{ss}}(y|x)$, where
\begin{align}
\pi_\mathrm{ss}(y|x)=\dfrac{1}{\sqrt{2\pi\bar{D}^Y}}\exp\left[-\dfrac{1}{2\bar{D}^Y}\left(y-\bar{\omega}^{YX}x\right)^2\right].
\label{ex2:pi}
\end{align}
The resulting effective dynamics for $X$ reads
\begin{align}
\dot{x}_\tau&=-\left(1-\bar{\omega}^{XY}\bar{\omega}^{YX}\right)x_\tau+\sqrt{2\bar{D}^X}\xi^X_\tau.
\label{ex2: effective dynamics}
\end{align}

\subsubsection{Trade-off between negative entropy production and efficiency}
We first consider the trade-off between the negative entropy production and efficiency (\ref{trade-off between entropy production and efficiency}).
Note that, unlike the previous example, this trade-off is not the same as the trade-off between power and efficiency (\ref{trade off power information-thermodynamic efficiency}) because there is no externally applied work in this system.
In the steady state with the fast relaxation limit of $Y$, the entropy production rate associated with $X$ reads
\begin{align}
\dot{S}^X_{\mathrm{env}}&=\dfrac{1}{D^X}\langle(-\omega^X x_t+\omega^{XY} y_t)\circ\dot{x}_t\rangle\notag\\
&=\omega^X\bar{\omega}^{XY}\bar{\omega}^{YX}\left(\dfrac{\bar{\omega}^{XY} \bar{D}^Y}{\bar{\omega}^{YX} \bar{D}^X}-1\right),
% &=\dfrac{\omega^{XY}\omega^{YX}}{\omega^Y}\left(\dfrac{\omega^{XY} D^Y}{\omega^{YX} D^X}-1\right).
\label{ex2:medium entropy}
\end{align}
where the symbol $\circ$ denotes the Stratonovich product.
We note that $\dot{S}^X_{\mathrm{env}}$ is induced by the fast variable $y_t$, which does not appear in the effective dynamics for $X$ (\ref{ex2: effective dynamics}).
In other words, $\dot{S}^X_{\mathrm{env}}$ is an entropy production invisible from the effective dynamics, which is called \textit{hidden entropy}~\cite{celani2012anomalous,kawaguchi2013fluctuation}.
Similarly, the information flow can be calculated as
\begin{align}
\dot{I}^X&=\int dxdyJ^X_{\mathrm{ss}}(x,y)\partial_x\ln\dfrac{p_{\mathrm{ss}}(x,y)}{p^X_{\mathrm{ss}}(x)p^Y_{\mathrm{ss}}(y)}\notag\\
% &=\int dxdy\left[(\omega^{XY} s-\omega^X x)p_{\mathrm{ss}}(x,y)-D^X\partial_xp_{\mathrm{ss}}(x,y)\right]\partial_x\ln\pi_{\mathrm{ss}}(y|x)\notag\\
% &=\omega^X\int dxdy\left[(\bar{\omega}^{XY} y-x)p_{\mathrm{ss}}(x,y)-\bar{D}^X\partial_xp_{\mathrm{ss}}(x,y)\right]\partial_x\ln\pi_{\mathrm{ss}}(y|x)\notag\\
% &=\omega^X\dfrac{\bar{\omega}^{YX}}{\bar{D}^Y}\left[\left\langle(\bar{\omega}^{XY} y-x)\left(y-\bar{\omega}^{YX}x\right)\right\rangle-\bar{\omega}^{YX} \bar{D}^X\right]\notag\\
% &=\dfrac{(\omega^{YX})^2D^X}{\omega^Y D^Y}\left(\dfrac{\omega^{XY} D^Y}{\omega^{YX} D^X}-1\right)\notag\\
&=\dfrac{\bar{\omega}^{YX} \bar{D}^X}{\bar{\omega}^{XY} \bar{D}^Y}\dot{S}^X_{\mathrm{env}}.
\label{ex2:LR}
\end{align}
In the context of the Brownian gyrator, we can show that there is a torque, which remains finite even in the fast relaxation limit of $Y$.
Both the medium entropy production rate $\dot{S}^X_{\mathrm{env}}$ and the information flow $\dot{I}^X$ are proportional to this ``hidden'' torque.
The fluctuation of the entropy production can be calculated by considering the tilted dynamics.
The result reads (see Appendix~\ref{Appendix: Coupled linear overdamped Langevin equations} for the derivation)
\begin{align}
D_S=\omega^X\dfrac{\bar{D}^Y}{\bar{D}^X}(\bar{\omega}^{XY})^2.
% &=\dfrac{1}{2}\left.\dfrac{\partial^2}{\partial \lambda^2}\mu(\lambda)\right|_{\lambda=0}\notag\\
% &=\dfrac{1}{2}\omega^X\left.\dfrac{\partial^2}{\partial \lambda^2}\theta_{\mathrm{max}}(\lambda)\right|_{\lambda=0}\notag\\
\label{ex2:D_S}
\end{align}

We now focus on the case where $\omega^{XY}\omega^{YX}>0$ and $\omega^{XY} D^Y<\omega^{YX} D^X$.
In this case, both the entropy production rate and information flow become negative, i.e., $X$ acts as an information-thermodynamic engine.
Then, the corresponding information-thermodynamic efficiency is given by
\begin{align}
\eta^X_S=\dfrac{|\dot{S}^X_{\mathrm{env}}|}{|\dot{I}^X|}=\dfrac{\omega^{XY} D^Y}{\omega^{YX} D^X}\le1.
\label{ex2:eta^X}
\end{align}
Combining (\ref{ex2:eta^X}) and (\ref{ex2:D_S}), we find that the upper bound on the negative entropy production rate (\ref{trade-off between entropy production and efficiency}) is
\begin{align}
D_S\dfrac{1-\eta^X_S}{\eta^X_S}&=\omega^X\dfrac{\bar{D}^Y}{\bar{D}^X}(\bar{\omega}^{XY})^2\dfrac{1-\dfrac{\bar{\omega}^{XY} \bar{D}^Y}{\bar{\omega}^{YX} \bar{D}^X}}{\dfrac{\bar{\omega}^{XY} \bar{D}^Y}{\bar{\omega}^{YX} \bar{D}^X}}\notag\\
&=|\dot{S}^X_{\mathrm{env}}|.
\label{ex2:trade-off between entropy production and efficiency}
\end{align}
Thus, the equality condition is satisfied even far from equilibrium in this case.
This is in contrast to the standard long-time TUR, where the equality is guaranteed only in the near-equilibrium limit.

We next consider the trade-off relation where the negative entropy production is bounded by the fluctuation of the time-integrated stochastic information flow $D_I$ (\ref{trade-off between entropy production and efficiency via D_I}).
The fluctuation of the information flow $D_I$ can also be calculated by using the tilted dynamics as
\begin{align}
D_I=\omega^X\dfrac{\bar{D}^X}{\bar{D}^Y}(\bar{\omega}^{YX})^2,
\label{ex2:D_I}
\end{align}
which satisfies $\sqrt{D_S/D_I}=\eta^X_S$.
Therefore, from (\ref{tightness of the bounds}), it follows that the upper bound of (\ref{trade-off between entropy production and efficiency via D_I}) is exactly the same as that of (\ref{trade-off between entropy production and efficiency}).
This implies that the trade-off between the information flow and efficiency (\ref{trade-off between information and efficiency}) also achieves the equality in this case:
\begin{align}
D_I(1-\eta^X_S)&=\omega^X\dfrac{\bar{D}^X}{\bar{D}^Y}(\bar{\omega}^{YX})^2\left(1-\dfrac{\bar{\omega}^{XY} \bar{D}^Y}{\bar{\omega}^{YX} \bar{D}^X}\right)\notag\\
&=|\dot{I}^X|.
\label{ex2:trade-off between information and efficiency}
\end{align}

\begin{figure}[t]
\center
\includegraphics[width=8.6cm]{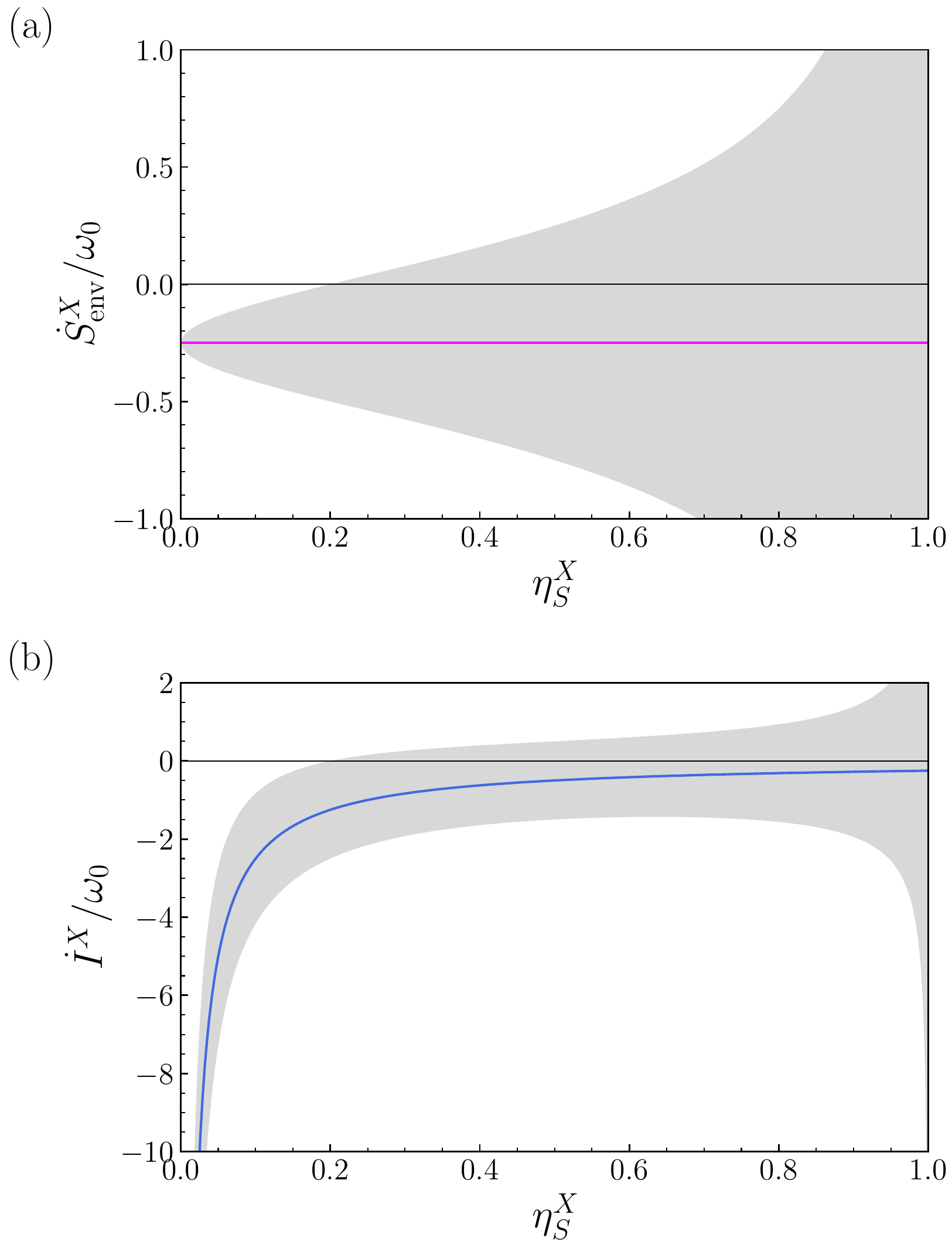}
\caption{$\eta^X_S$-dependence of the entropy production rate (a) and information flow (b) with the scaling $\omega^X=\omega_0/(1-\eta^X_S)$. Here, we assume that $\omega_0$ is constant. The gray shaded region represents the fluctuations of the entropy production and information flow quantified by $\sqrt{D_S/\omega_0}$ and $\sqrt{D_I/\omega_0}$, respectively. The parameter values are $\bar{\omega}^{XY}=\bar{\omega}^{XY}=0.5$.}
\label{fig:ex2_result}
\end{figure}

We now consider the possibility of achieving finite negative entropy production even when $\eta^X_S\rightarrow1$. 
We first note that the negative entropy production can be expressed in terms of $\eta^X_S$ as
\begin{align}
|\dot{S}^X_{\mathrm{env}}|=\omega^X\bar{\omega}^{XY}\bar{\omega}^{YX}\left(1-\eta^X_S\right).
% \label{ex2:medium entropy}
\end{align}
Since $0<\bar{\omega}^{XY}\bar{\omega}^{YX}<1$, we find that $|\dot{S}^X_{\mathrm{env}}|\rightarrow0$ as $\eta^X_S\rightarrow1$ as long as $\omega^X$ is finite.
In contrast, if $\omega^X$ is scaled as $\omega^X=\omega_0/(1-\eta^X_S)$, the negative entropy production can remain finite even in the limit $\eta^X_S\rightarrow1$:
\begin{align}
|\dot{S}^X_{\mathrm{env}}|=\omega_0\bar{\omega}^{XY}\bar{\omega}^{YX}.
% \label{ex2:medium entropy}
\end{align}
As can be seen from the trade-off relation (\ref{ex2:trade-off between entropy production and efficiency}), the fluctuation of entropy production blows up as $\eta^X_S\rightarrow1$ in this case (see Fig.~\ref{fig:ex2_result}(a)):
\begin{align}
D_S&=\omega^X\dfrac{\bar{D}^Y}{\bar{D}^X}(\bar{\omega}^{XY})^2\notag\\
&=\omega_0\dfrac{\eta^X_S}{1-\eta^X_S}\bar{\omega}^{XY}\bar{\omega}^{YX}.
% \label{ex2:D_S}
\end{align}
Similarly, the information flow can also remain finite in the limit $\eta^X_S\rightarrow1$
\begin{align}
|\dot{I}^X|=\omega_0\bar{\omega}^{XY}\bar{\omega}^{YX}\dfrac{1}{\eta^X_S},
\end{align}
at the expense of the blow-up of the fluctuation of information flow as $\eta^X_S\rightarrow1$ (see Fig.~\ref{fig:ex2_result}(b)):
\begin{align}
D_I&=\omega^X\dfrac{\bar{D}^X}{\bar{D}^Y}(\bar{\omega}^{YX})^2\notag\\
&=\omega_0\dfrac{1}{\eta^X_S(1-\eta^X_S)}\bar{\omega}^{XY}\bar{\omega}^{YX}.
\end{align}

\subsubsection{Equality condition of bipartite TUR for this model}
Here, we discuss the reason why the equality of the trade-offs (\ref{ex2:trade-off between entropy production and efficiency}) and (\ref{ex2:trade-off between information and efficiency}) is achieved in this model.
We first recall that these trade-offs are special cases of the bipartite TUR in the fast relaxation limit of $Y$ (\ref{bipartite TUR_fast relaxation limit}).
In this model, the time-integrated generalized current $\hat{\mathcal{J}}_{\mathcal{T}}$ for the subsystem $X$ can be expressed as
\begin{align}
\hat{\mathcal{J}}_{\mathcal{T}}&=\int^{t=\mathcal{T}}_{t=0}g(x_t,y_t)\circ dx_t
\end{align}
with an arbitrary weight function $g(x,y)$.
As in Sec.~\ref{Equality condition}, the current can be decomposed as $\hat{\mathcal{J}}_{\mathcal{T}}=\hat{\mathcal{J}}^{\mathrm{I}}_{\mathcal{T}}+\hat{\mathcal{J}}^{\mathrm{II}}_{\mathcal{T}}$ with
\begin{align}
\hat{\mathcal{J}}^{\mathrm{I}}_{\mathcal{T}}&:=\int^{t=\mathcal{T}}_{t=0}g(x_t,y_t)\cdot \sqrt{2D^X}dW^X_t,\\
\hat{\mathcal{J}}^{\mathrm{II}}_{\mathcal{T}}&:=\int^{\mathcal{T}}_0f(x_t,y_t)dt,
\end{align}
where the symbol $\cdot$ denotes the Ito product, $W^X_t$ denotes the Wiener process, and
\begin{align}
f(x,y):=g(x,y)(-\omega^X x+\omega^{XY} y)+D^X\partial_xg(x,y).
\end{align}

We now show that this model satisfies the sufficient condition for the bipartite TUR in the fast relaxation limit of $Y$ (\ref{bipartite TUR_fast relaxation limit}) to hold with equality, described in Sec.~\ref{Equality condition}.
First, the weight of the current should be proportional to that of the partial entropy production:
\begin{align}
g(x,y)&=C\dfrac{1}{2D^X}\dfrac{J^X_{\mathrm{ss}}(x_t,y_t)}{p_{\mathrm{ss}}(x_t,y_t)}\notag\\
&=C\dfrac{1}{2D^X}\omega^X\bar{\omega}^{XY}\left(1-\dfrac{\bar{\omega}^{YX} \bar{D}^X}{\bar{\omega}^{XY} \bar{D}^Y}\right)\left(y-\bar{\omega}^{YX}x\right).
\label{ex2: condition}
\end{align}
The time-integrated stochastic information flow $\Delta\hat{I}^X$ in the steady state with the fast relaxation limit of $Y$ is an example that satisfies this condition:
\begin{align}
\Delta\hat{I}^X&=\int^{t=\mathcal{T}}_{t=0}\partial_x\ln\dfrac{p_{\mathrm{ss}}(x_t,y_t)}{p^X_{\mathrm{ss}}(x_t)p^Y_{\mathrm{ss}}(y_t)}\circ dx_t\notag\\
&\simeq\int^{t=\mathcal{T}}_{t=0}\partial_x\ln\pi_{\mathrm{ss}}(y_t|x_t)\circ dx_t\notag\\
&=\int^{t=\mathcal{T}}_{t=0}\dfrac{\bar{\omega}^{YX}}{\bar{D}^Y}(y_t-\bar{\omega}^{YX}x_t)\circ dx_t.
% &=\dfrac{\bar{\omega}^{YX}}{\bar{D}^Y}\int^{\tau=\omega^X\mathcal{T}}_{\tau=0}y_\tau\circ dx_\tau-\left.\dfrac{(\bar{\omega}^{YX})^2}{2\bar{D}^Y}x^2_\tau\right|^{\tau=\omega^X\mathcal{T}}_{\tau=0}.
\end{align}
Second, for this choice of the current, the fluctuation of $\hat{\mathcal{J}}^{\mathrm{II}}_{\mathcal{T}}$ must go to zero in the fast relaxation limit of $Y$:
\begin{align}
D^{\mathrm{II}}_{\mathcal{J}}:=\lim_{\mathcal{T}\rightarrow\infty}\dfrac{1}{2\mathcal{T}}\mathrm{Var}[\hat{\mathcal{J}}^{\mathrm{II}}_{\mathcal{T}}]=0.
\end{align}
In this model, we can confirm that this condition is indeed satisfied by explicitly calculating $D^{\mathrm{II}}_{\mathcal{J}}$ (see Appendix~\ref{Proof for linear weight function}).
As a result, the equality of (\ref{bipartite TUR_fast relaxation limit}) is achieved for a current that satisfies the condition (\ref{ex2: condition}) in the fast relaxation limit of $Y$:
\begin{align}
\dfrac{D_{\mathcal{J}}}{J^2}=\dfrac{1}{\dot{S}^X_{\mathrm{tot}}-\dot{I}^X}.
\label{ex2: equality of bipartite TUR}
\end{align}

Note that the condition described above is only a sufficient condition.
In fact, the equality (\ref{ex2: equality of bipartite TUR}) holds for more diverse types of currents that do not even satisfy the condition (\ref{ex2: condition}) in this model.
To see this, note that the current $\hat{\mathcal{J}}_{\mathcal{T}}$ that satisfies the condition (\ref{ex2: condition}) can be expressed as
\begin{align}
\hat{\mathcal{J}}_{\mathcal{T}}&=C\int^{t=\mathcal{T}}_{t=0}\left(y_t-\bar{\omega}^{YX}x_t\right)\circ dx_t\notag\\
&=C\left[\int^{t=\mathcal{T}}_{t=0}y_t\circ dx_t-\left.\dfrac{1}{2}\bar{\omega}^{YX}x^2_t\right|^{t=\mathcal{T}}_{t=0}\right].
\label{ex2: J_boundary term}
\end{align}
The important point here is that the second term is a boundary term and can be ignored when considering the long-time statistical properties of $\hat{\mathcal{J}}_{\mathcal{T}}$.
(For the effect of such a boundary term on the large deviation, see~\cite{du2023dynamical}.)
Therefore, any current $\hat{\mathcal{J}}_{\mathcal{T}}$ that has the same long-time statistical properties as (\ref{ex2: J_boundary term}) satisfies the equality (\ref{ex2: equality of bipartite TUR}).
The example includes the stochastic medium entropy production $\Delta\hat{S}^X_{\mathrm{env}}$:
\begin{align}
\Delta\hat{S}^X_{\mathrm{env}}&=\int^{t=\mathcal{T}}_{t=0}\dfrac{1}{D^X}(-\omega^X x_t+\omega^{XY} y_t)\circ dx_t\notag\\
% &=\int^{\tau=\omega^X\mathcal{T}}_{\tau=0}\dfrac{1}{\bar{D}^X}(-x_\tau+\bar{\omega}^{XY} y_\tau)\circ dx_\tau\notag\\
&=\dfrac{\bar{\omega}^{XY}}{\bar{D}^X}\int^{t=\mathcal{T}}_{t=0}y_t\circ dx_t-\left.\dfrac{1}{2\bar{D}^X}x^2_t\right|^{t=\mathcal{T}}_{t=0}.
\end{align}
Hence, the choice $\hat{\mathcal{J}}_{\mathcal{T}}=\Delta\hat{S}^X_{\mathrm{env}}$ also satisfies the equality (\ref{ex2: equality of bipartite TUR}), although it does not satisfy the condition (\ref{ex2: condition}).
Indeed, we can show that the following relation holds:
\begin{align}
\dfrac{D^{\mathrm{I}}_S}{(\dot{S}^X_{\mathrm{env}})^2}>\dfrac{D_S}{(\dot{S}^X_{\mathrm{env}})^2}=\dfrac{1}{\dot{S}^X_{\mathrm{tot}}-\dot{I}^X}.
\end{align}
Here, $D^{\mathrm{I}}_S:=\lim_{\mathcal{T}\rightarrow\infty}\mathrm{Var}[\hat{\mathcal{J}}^{\mathrm{I}}_{\mathcal{T}}]/2\mathcal{T}$ denotes the fluctuation of $\hat{\mathcal{J}}^{\mathrm{I}}_{\mathcal{T}}$ with $\hat{\mathcal{J}}_{\mathcal{T}}=\Delta\hat{S}^X_{\mathrm{env}}$, which is given by
\begin{align}
D^{\mathrm{I}}_S=\omega^X\left[\dfrac{\bar{D}^Y}{\bar{D}^X}(\bar{\omega}^{XY})^2+(1-\bar{\omega}^{XY}\bar{\omega}^{YX})\right].
\end{align}

\subsubsection{Input-output fluctuation inequalities}
We finally consider the input-output fluctuation inequalities for the case of $\omega^{XY}\omega^{YX}>0$ and $\omega^{XY} D^Y<\omega^{YX} D^X$, where the entropy production and the information flow correspond to the output and input currents, respectively.
From the relation $\sqrt{D_S/D_I}=\eta^X_S$, it immediately follows that $D_S\le D_I$ and $D_I/(\dot{I}^X)^2=D_S/(\dot{S}^X_{\mathrm{env}})^2$.
Thus, as in the previous example, the input-output fluctuation inequalities are satisfied even beyond the linear response regime in this model, and the equality is achieved for the inequality regarding the relative fluctuations.

\section{Concluding remarks\label{Concluding remarks}}
%%%%%% Summary
In this paper, we have obtained several fundamental limits for information processing systems.
Specifically, we have derived a TUR-type inequality for bipartite systems that provides a universal lower bound on the relative fluctuation of an arbitrary current for a system of interest by the associated partial entropy production, which includes the information flow.
This bipartite TUR includes the standard TUR as a special case and incorporates the effect of the interaction with external auxiliary systems.
As a corollary to this inequality, we have derived universal trade-off relations between the negative entropy production rate and the information-thermodynamic efficiency, which can be regarded as an extension of the trade-offs for heat engines~\cite{shiraishi2016universal,pietzonka2018universal} to information-thermodynamic engines.
Furthermore, in the fast relaxation limit of the auxiliary system, we have shown that the Gallavotti-Cohen symmetry holds even in the presence of information flow.
From this symmetry, we can show that the input-output fluctuation inequalities are also valid for information processing systems.
We have illustrated our results with two simple examples: coupled quantum dots and coupled linear overdamped Langevin equations.
In particular, we have seen that the latter provides an example where the equality of the bipartite TUR is achieved even far from equilibrium.

%%%%%% Remark
Here, we provide some remarks on previous studies related to our results.
We first note that the bipartite TUR in the short time limit $\mathcal{T}\rightarrow0$ is already proved in~\cite{otsubo2020estimating} using the Cauchy-Schwarz inequality.
Our first main result (\ref{bipartite TUR}) can be regarded as an extension of the short-time bipartite TUR to an arbitrary observation time $\mathcal{T}$.
TUR-type inequalities including measurement and feedback are also derived from fluctuation theorems in~\cite{potts2019thermodynamic,van2020uncertainty}.
While these relations include a contribution of information induced by measurement and feedback processes, this contribution appears in the form of total entropy production rather than partial entropy production.
Therefore, our bipartite TUR can provide more stringent bounds on the precision of currents under measurement and feedback control.
The standard TUR has also been discussed as a tool for inferring entropy production~\cite{li2019quantifying,vu2020entropy,otsubo2020estimating,manikandan2020inferring,otsubo2022estimating,das2022inferring}.
In this context, the bipartite TUR proved here may provide a promising approach to estimating a partial entropy production, especially an information flow.

Next, we remark on the range of validity of the bipartite TUR.
While here we have presented the bipartite TUR in the steady state, this relation is valid even for systems under arbitrary time-dependent driving from arbitrary initial states.
In Appendix~\ref{Bipartite TUR for overdamped Langevin equations}, we provide a proof of the bipartite TUR in a general form for the case of overdamped Langevin equations. 
It should also be noted that the bipartite TUR is generally not valid for systems with broken time-reversal symmetry, such as underdamped Langevin dynamics~\cite{van2019uncertainty,lee2019thermodynamic,fischer2020free,lee2021universal,fu2022thermodynamic,dechant2022bounds,pietzonka2022classical}, as in the standard TUR.
However, many relevant biological systems are often described by continuous-time Markov jump processes or diffusion processes with only even variables and parameters under time reversal.
Therefore, the results described in this paper will be applicable to a wide range of systems, including biological systems.

In this study, we have focused mainly on the case where an auxiliary system evolves much faster than the system of interest.
Such a separation of time scales allows the dynamics of a composite system to be reduced to the effective dynamics of the system of interest, and thus various universal relations similar to those found for a single system hold.
While we expect such a separation of time scales to be ubiquitous in biological systems, extending our results to cases where there is no clear time-scale separation would be important for elucidating the design principles of biological systems.

\begin{acknowledgements}
We are grateful to Takayuki Ariga for the fruitful discussion.
TT is supported by JSPS KAKENHI Grant Number JP23K19035 and JST, PRESTO Grant Number JPMJPR23O6, Japan.
TVV is supported by JSPS KAKENHI Grant Number JP23K13032.
KS is supported by JSPS KAKENHI Grant Numbers JP23H01099, JP19H05603, and JP19H05791.
\end{acknowledgements}

\appendix
\section{Bipartite TUR for overdamped Langevin equations\label{Bipartite TUR for overdamped Langevin equations}}
In this section, we derive the bipartite TUR for overdamped Langevin equations.
While the derivation based on the generalized Cram\'er-Rao inequality described in Sec.~\ref{Derivation of the bipartite TUR} is also valid for this case, here we prove the bipartite TUR more directly from the Langevin equations, following Ref.~\cite{dieball2023direct}.
We provide a proof of the bipartite TUR in a general form that is valid not only for a steady state but also for systems under arbitrary time-dependent driving from arbitrary initial states.
We note that this direct approach is valid even for Markov jump processes~\cite{dieball2023direct}.

We consider the following coupled overdamped Langevin equations:
\begin{align}
\dot{x}_t&=F^X_t(x_t,y_t)+\sqrt{2D^X}\xi^X_t,\label{Langevin-X}\\
\dot{y}_t&=F^Y_t(x_t,y_t)+\sqrt{2D^Y}\xi^Y_t,\label{Langevin-Y}
\end{align}
where $F^Z_t(x,y)$ ($Z=X,Y$) denotes the time-dependent drift term, $D^Z$ denotes the noise intensity, and $\xi^Z_t$ is a zero-mean white Gaussian noise that satisfies $\langle\xi^Z_t\xi^{Z'}_{t'}\rangle=\delta_{ZZ'}\delta(t-t')$.
The independence of the noises $\xi^X_t$ and $\xi^Y_t$ ensures that the system satisfies the bipartite property.
The corresponding Fokker-Planck equation reads
\begin{align}
\partial_tp_t(x,y)=-\partial_xJ^X_t(x,y)-\partial_yJ^Y_t(x,y),
\label{FP}
\end{align}
where $J^Z_t$ denotes the probability current:
\begin{align}
J^X_t(x,y)&:=F^X_t(x,y)p_t(x,y)-D^X\partial_xp_t(x,y),\\
J^Y_t(x,y)&:=F^Y_t(x,y)p_t(x,y)-D^Y\partial_yp_t(x,y).
\end{align}
Let $\hat{\mathcal{J}}_\mathcal{T}$ be the time-integrated generalized current for the subsystem $X$ with an arbitrary time-dependent weight function $g_t(x,y)$:
\begin{align}
\hat{\mathcal{J}}_{\mathcal{T}}:=\int^{t=\mathcal{T}}_{t=0}g_t(x_t,y_t)\circ dx_t.
\end{align}
Converting from the Stratonovich to the Ito product, the current can be decomposed into two parts $\hat{\mathcal{J}}_{\mathcal{T}}=\hat{\mathcal{J}}^{\mathrm{I}}_{\mathcal{T}}+\hat{\mathcal{J}}^{\mathrm{II}}_{\mathcal{T}}$ with
\begin{align}
\hat{\mathcal{J}}^{\mathrm{I}}_{\mathcal{T}}&:=\int^{t=\mathcal{T}}_{t=0}g_t(x_t,y_t)\cdot \sqrt{2D^X}dW^X_t,\\
\hat{\mathcal{J}}^{\mathrm{II}}_{\mathcal{T}}&:=\int^{\mathcal{T}}_0f_t(x_t,y_t)dt,
\end{align}
where $W^X_t$ denotes the Wiener process, and 
\begin{align}
f_t(x,y):=g_t(x,y)F^X_t(x,y)+D^X\partial_xg_t(x,y).
\end{align}

We introduce the following quantity:
\begin{align}
\hat{A}_{\mathcal{T}}:=\int^{t=\mathcal{T}}_{t=0}\dfrac{1}{2D^X}\dfrac{J^X_t(x_t,y_t)}{p_t(x_t,y_t)}\cdot\sqrt{2D^X}dW^X_t.
\end{align}
The second moment of this quantity gives the partial entropy production for $X$:
\begin{align}
\langle\hat{A}^2_{\mathcal{T}}\rangle&=\int^{\mathcal{T}}_0dt\int dxdy\dfrac{1}{2D^X}\dfrac{[J^X_t(x,y)]^2}{p_t(x,y)}\notag\\
&=\dfrac{1}{2}\Delta\sigma^X=\dfrac{1}{2}(\Delta S^X_{\mathrm{tot}}-\Delta I^X).
\end{align}
Furthermore, we can easily confirm that $\langle\hat{A}_{\mathcal{T}}\rangle=0$ and $\langle\hat{A}_{\mathcal{T}}\hat{\mathcal{J}}^{\mathrm{I}}_{\mathcal{T}}\rangle=\langle\hat{\mathcal{J}}_{\mathcal{T}}\rangle$.
Therefore, we find that
\begin{align}
\langle\hat{A}_{\mathcal{T}}(\hat{\mathcal{J}}_{\mathcal{T}}-\langle\hat{\mathcal{J}}_{\mathcal{T}}\rangle)\rangle=\langle\hat{\mathcal{J}}_{\mathcal{T}}\rangle+\langle\hat{A}_{\mathcal{T}}\hat{\mathcal{J}}^{\mathrm{II}}_{\mathcal{T}}\rangle.
\end{align}
By using the Cauchy-Schwarz inequality, we obtain
\begin{align}
[\langle\hat{\mathcal{J}}_{\mathcal{T}}\rangle+\langle\hat{A}_{\mathcal{T}}\hat{\mathcal{J}}^{\mathrm{II}}_{\mathcal{T}}\rangle]^2&=\langle\hat{A}_{\mathcal{T}}(\hat{\mathcal{J}}_{\mathcal{T}}-\langle\hat{\mathcal{J}}_{\mathcal{T}}\rangle)\rangle^2\notag\\
&\le\langle\hat{A}^2_{\mathcal{T}}\rangle\mathrm{Var}[\hat{\mathcal{J}}_{\mathcal{T}}]\notag\\
&=\dfrac{1}{2}(\Delta S^X_{\mathrm{tot}}-\Delta I^X)\mathrm{Var}[\hat{\mathcal{J}}_{\mathcal{T}}],
\label{Bipartite TUR_direct route}
\end{align}
which has a form similar to that of the bipartite TUR (\ref{bipartite TUR}).
In fact, we can show that the additional current term $\langle\hat{A}_{\mathcal{T}}\hat{\mathcal{J}}^{\mathrm{II}}_{\mathcal{T}}\rangle$ in (\ref{Bipartite TUR_direct route}) exactly equal to $\langle\hat{\mathcal{J}}_{\mathcal{T}}\rangle_q$ in the bipartite TUR (\ref{bipartite TUR}).
To see this, we first rewrite $\langle\hat{A}_{\mathcal{T}}\hat{\mathcal{J}}^{\mathrm{II}}_{\mathcal{T}}\rangle$ as follows~\cite{dieball2023direct}:
\begin{widetext}
\begin{align}
\langle A_{\mathcal{T}}J^{\mathrm{II}}_{\mathcal{T}}\rangle&=\left\langle\int^{t'=\mathcal{T}}_{t'=0}\dfrac{1}{2D^X}\dfrac{J^X_{t'}(x_{t'},y_{t'})}{p_{t'}(x_{t'},y_{t'})}\cdot\sqrt{2D^X}dW^X_{t'}\int^{\mathcal{T}}_0f_t(x_t,y_t)dt\right\rangle\notag\\
&=-\int^{\mathcal{T}}_0dt\int dxdyf_t(x,y)\int^t_0dt'\int dx'dy'p(x,y,t|x',y',t')\partial_{x'}J^X_{t'}(x',y')\notag\\
&=\int^{\mathcal{T}}_0dt\int dxdyg_t(x,y)[F^X_t(x,y)-D^X\partial_x]\tilde{q}_t(x,y),
\label{AJ in terms of q_tilde}
\end{align}
where
\begin{align}
\tilde{q}_t(x,y):=-\int^t_0dt'\int dx'dy'p(x,y,t|x',y',t')\partial_{x'}J^X_{t'}(x',y').
\label{q_tilde}
\end{align}
In the second line of (\ref{AJ in terms of q_tilde}), we have used the Doob transform~\cite{doob1957conditional,chetrite2015nonequilibrium,majumdar2015effective,pigolotti2017generic,dechant2021continuous}, which maps a stochastic process conditioned on a future event (in this case, $(x_t,y_t)=(x,y)$) to an unconditioned stochastic process with an additional drift term.
By differentiating (\ref{q_tilde}) with respect to $t$, we obtain the time evolution equation of $\tilde{q}_t(x,y)$:
\begin{align}
\partial_t\tilde{q}_t(x,y)&=-\int dx'dy'\delta(x-x')\delta(y-y')\partial_{x'}J^X_t(x',y')-\int^t_0dt'\int dx'dy'\partial_tp(x,y,t|x',y',t')\partial_{x'}J^X_{t'}(x',y')\notag\\
&=-\partial_xJ^X_t(x,y)+\mathcal{L}_t[\tilde{q}](x,y),
\end{align}
\end{widetext}
with $\tilde{q}_0=0$, where $\mathcal{L}_t$ denotes the Fokker-Planck operator.
Thus, $\langle A_{\mathcal{T}}J^{\mathrm{II}}_{\mathcal{T}}\rangle$ is exactly equal to $\langle\hat{\mathcal{J}}_{\mathcal{T}}\rangle_q$.
We remark that this conclusion is also confirmed by noting that $\hat{A}_{\mathcal{T}}$ corresponds to the $\theta$-derivative of the path probability $\partial_\theta\ln\mathbb{P}_\theta(\Gamma)|_{\theta=0}$ used in the generalized Cram\'er-Rao inequality.

Note that $\langle\hat{\mathcal{J}}_{\mathcal{T}}\rangle_q$ generally reflects not only the contribution of interaction with $Y$ but also the effect of nonstationarity.
This point will be clarified in the next section.

\section{Relation to the conventional transient TUR\label{Appendix: Transient TUR}}
In this section, we consider the bipartite TUR in a general form that is applicable to a transient state, derived in the previous section.
Here, we prove that $\langle \hat{\mathcal{J}}_\mathcal{T}\rangle_q=\mathcal{T}\partial_{\mathcal{T}}\langle \hat{\mathcal{J}}_\mathcal{T}\rangle-\langle \hat{\mathcal{J}}_\mathcal{T}\rangle$, if the system is time-homogeneous and the transition rate for $X$ and the weight are independent of $Y$ as $w^y_{xx'}=w_{xx'}$ and $d^y_{xx'}=d_{xx'}$.
In this case, the bipartite TUR becomes
\begin{align}
\dfrac{\mathrm{Var}[\hat{\mathcal{J}}_\mathcal{T}]}{\left[\mathcal{T}\partial_{\mathcal{T}}\langle \hat{\mathcal{J}}_\mathcal{T}\rangle\right]^2}\ge\dfrac{2}{\Delta S^X_{\mathrm{tot}}-\Delta I^X},
\label{Transient TUR}
\end{align}
which has a form similar to the conventional transient TUR~\cite{liu2020thermodynamic,dieball2023direct}.
From this result, we can also confirm that $\langle \hat{\mathcal{J}}_\mathcal{T}\rangle_q=0$ when the system is in the steady state.
While we focus on the Markov jump processes in the following, the same result can be obtained for diffusion processes.

By noting that the transition rate $w^y_{xx'}=w_{xx'}$ and the weight of the current $d^y_{xx'}=d_{xx'}$ do not depend on $Y$, we obtain
\begin{align}
\langle \hat{\mathcal{J}}_\mathcal{T}\rangle_q&=\int^\mathcal{T}_0dt\sum_{x}\sum_{x'(\neq x)}\sum_yw_{xx'}q_t(x',y)d_{xx'}\notag\\
&=\int^\mathcal{T}_0dt\sum_{x}\sum_{x'(\neq x)}w_{xx'}q^X_t(x')d_{xx'}.
\label{Jq_q^X}
\end{align}
As in (\ref{q_tilde}), we can easily show that $q^X_t(x')=\sum_yq_t(x,y)$ has the form
\begin{align}
q^X_t(x)&=\int^t_0dt'\sum_{x'}p(x,t|x',t')\sum_{x''}w_{x'x''}p^X_{t'}(x'')\notag\\
&=\int^t_0dt'\sum_{x'}p(x,t|x',t')\partial_{t'}p^X_{t'}(x').
\label{q^X}
\end{align}
By substituting (\ref{q^X}) into (\ref{Jq_q^X}) and integrating by parts, we obtain
\begin{align}
\langle \hat{\mathcal{J}}_\mathcal{T}\rangle_q&=\int^\mathcal{T}_0dt\sum_{x}\sum_{x'(\neq x)}w_{xx'}d_{xx'}\notag\\
&\quad\times\int^t_0dt'\sum_{x'}p(x,t|x',t')\partial_{t'}p^X_{t'}(x')\notag\\
&=-\int^\mathcal{T}_0dt\sum_{x}\sum_{x'(\neq x)}w_{xx'}d_{xx'}\notag\\
&\quad\times\int^t_0dt'\sum_{x'}\partial_{t'}p(x,t|x',t')p^X_{t'}(x').
\end{align}
Since the system is time-homogeneous, we have $\partial_{t'}p(x,t|x',t')=-\partial_tp(x,t|x',t')$, and thus
\begin{align}
&\quad\int^t_0dt'\sum_{x'}\partial_{t'}p(x,t|x',t')p^X_{t'}(x')\notag\\
&=-\int^t_0dt'\partial_t\sum_{x'}p(x,t|x',t')p^X_{t'}(x')\notag\\
&=-t\partial_tp^X_t(x).
\end{align}
Hence, by integrating by parts, we obtain
\begin{align}
\langle \hat{\mathcal{J}}_\mathcal{T}\rangle_q&=\int^\mathcal{T}_0dt\sum_{x}\sum_{x'(\neq x)}w_{xx'}d_{xx'}t\partial_tp^X_t(x)\notag\\
&=\mathcal{T}\partial_{\mathcal{T}}\langle \hat{\mathcal{J}}_\mathcal{T}\rangle-\langle \hat{\mathcal{J}}_\mathcal{T}\rangle.
\end{align}

\section{Coupled quantum dots\label{Appendix: Coupled quantum dots}}
In this section, we provide a detailed calculation of the fluctuation of the chemical work $D_W$ in the fast relaxation limit of $Y$ for the coupled quantum dots introduced in Sec.~\ref{Coupled quantum-dots}.
The fluctuation of the information flow $D_I$ can be calculated in a similar way.
The stochastic chemical work is defined as
\begin{align}
\Delta\hat{W}^X:=\sum_\nu\sum_{x}\sum_{x'(\neq x)}\sum_y\hat{n}^{(\nu)y}_{xx'}b_{xx'}\mu_\nu,
\end{align} 
where
\begin{align}
b_{xx'}:=
\begin{cases}
1\quad&(x=1, x'=0),\\
-1\quad&(x=0, x'=1).
\end{cases}
\end{align}
Then, the fluctuation of the stochastic chemical work is defined as
\begin{align}
D_W:=\lim_{\mathcal{T}\rightarrow\infty}\dfrac{1}{2\mathcal{T}}\mathrm{Var}[\Delta\hat{W}^X].
\end{align}
The fluctuation $D_W$ can be obtained from the scaled cumulant generating function defined by
\begin{align}
\mu(\lambda)&=\lim_{\mathcal{T}\rightarrow\infty}\dfrac{1}{\mathcal{T}}\ln\langle e^{\lambda\Delta\hat{W}^X}\rangle.
% &=\lim_{\mathcal{T}\rightarrow\infty}\dfrac{1}{\mathcal{T}}\ln\sum_x\sum_yG_{\mathcal{T}}(x,y),
\end{align}
As described in Sec.~\ref{Gallavotti-Cohen symmetry}, the scaled cumulant generating function can be calculated by considering the generating function conditioned to a final state $(x,y)$:
\begin{align}
G_{\mathcal{T}}(x,y)&:=\int d\Delta W^Xp_\mathcal{T}(x,y,\Delta W^X)e^{\lambda\Delta W^X}.
\end{align}
The time evolution of $G_{\mathcal{T}}(x,y)$ reads
\begin{align}
\partial_{\tau}G_{\tau}(x,y)&=\sum_\nu\sum_{x'}\left[\mathcal{L}^X_\lambda\right]^{(\nu)y}_{xx'}G_{\tau}(x',y)\notag\\
&\qquad+\dfrac{1}{\epsilon}\sum_{y'}\left[\mathcal{L}^Y_\lambda\right]^{yy'}_xG_{\tau}(x,y'),
% \partial_{\tau}G_{\tau}(x,y)&=\sum_{x'(\neq x)}\left[\widetilde{w}^y_{xx'}\exp\left(\lambda\ln\dfrac{\widetilde{w}^y_{xx'}}{\widetilde{w}^y_{x'x}}-\lambda_I\ln\dfrac{\pi_{\mathrm{ss}}(y|x)}{\pi_{\mathrm{ss}}(y|x')}\right)G_{\tau}(x',y)-\widetilde{w}^y_{x'x}G_{\tau}(x,y)\right]\notag\\
% &\qquad+\dfrac{1}{\epsilon}\sum_{y'(\neq y)}\left[\widetilde{w}^{yy'}_xG_{\tau}(x,y')-\widetilde{w}^{y'y}_xG_{\tau}(x,y)\right].
\label{ex1: time evolution of G}
\end{align}
where $\mathcal{L}^X_\lambda$ and $\mathcal{L}^Y_\lambda$ denote the tilted generators given by
\begin{align}
\left[\mathcal{L}^X_\lambda\right]^{(\nu)y}_{xx'}:=
\begin{cases}
\widetilde{w}^{(\nu)y}_{xx'}e^{\lambda b_{xx'}\mu_\nu}\quad&(x\neq x'),\\
-\sum_{x'(\neq x)}\widetilde{w}^{(\nu)y}_{x'x}\quad&(x=x'),
\end{cases}
\end{align}
\begin{align}
\left[\mathcal{L}^Y_\lambda\right]^{yy'}_x:=
\begin{cases}
\widetilde{w}^{yy'}_x\quad&(y\neq y'),\\
-\sum_{y'(\neq y)}\widetilde{w}^{y'y}_x\quad&(y=y'),
\end{cases}
\end{align}
where we have used the dimensionless slow time $\tau:=\Gamma_X\mathcal{T}$ and dimensionless transition rates $\widetilde{w}^{(\nu)y}_{xx'}:=w^{(\nu)y}_{xx'}/\Gamma_X$ and $\widetilde{w}^{yy'}_x:=w^{yy'}_x/\Gamma_Y$ with a small parameter $\epsilon:=\Gamma_X/\Gamma_Y\ll1$ (do not confuse $\epsilon$ with the error probability $\varepsilon$).

Since we are interested in the fast relaxation limit of $Y$, we can consider the effective tilted dynamics for $G^X_\tau:=\sum_yG_\tau$.
By performing a perturbation expansion as in Sec.~\ref{Gallavotti-Cohen symmetry}, we obtain
\begin{align}
\partial_\tau G^X_{\tau}(x)=\sum_{x'}\left[\overline{\mathcal{L}}^X_\lambda\right]_{xx'}G^X_\tau(x'),
% \partial_\tau G^X_{\tau}(x)=\sum_{x'(\neq x)}\left\{\sum_y\left[\widetilde{w}^y_{xx'}\exp\left(\lambda_S\ln\dfrac{\widetilde{w}^y_{xx'}}{\widetilde{w}^y_{x'x}}-\lambda_I\ln\dfrac{\pi_{\mathrm{ss}}(y|x)}{\pi_{\mathrm{ss}}(y|x')}\right)\pi_{\mathrm{ss}}(y|x')\right]G^X_\tau(x')-\sum_y\left[\widetilde{w}^y_{x'x}\pi_{\mathrm{ss}}(y|x)\right]G^X_\tau(x)\right\}.
\end{align}
where $\overline{\mathcal{L}}^X_\lambda$ denotes the effective tilted generator given by
\begin{align}
\left[\overline{\mathcal{L}}^X_\lambda\right]_{xx'}&:=\sum_\nu\sum_y\left[\mathcal{L}^X_\lambda\right]^{(\nu)y}_{xx'}\pi_{\mathrm{ss}}(y|x'),
% &=
% \begin{cases}
% \sum_\nu\sum_y\left[\widetilde{w}^{(\nu)y}_{xx'}\exp\left(\lambda\ln\dfrac{\widetilde{w}^{(\nu)y}_{xx'}}{\widetilde{w}^{(\nu)y}_{x'x}}\right)\pi_{\mathrm{ss}}(y|x')\right]\quad&(x\neq x'),\\
% -\sum_\nu\sum_{x'(\neq x)}\left[\sum_y\widetilde{w}^{(\nu)y}_{x'x}\pi_{\mathrm{ss}}(y|x)\right]\quad&(x=x').
% \end{cases}
\end{align}
which can be expressed as
\begin{align}
\overline{\mathcal{L}}^X_\lambda=
\begin{pmatrix}
-\sum_\nu \overline{w}^{(\nu)}_{10} & \sum_\nu\overline{w}^{(\nu)}_{01}e^{-\lambda\mu_\nu} \\
\sum_\nu\overline{w}^{(\nu)}_{10}e^{\lambda\mu_\nu} & -\sum_\nu \overline{w}^{(\nu)}_{01}
\end{pmatrix}
,
\end{align}
where we have introduced the effective transition rate $\overline{w}^{(\nu)}_{xx'}:=\sum_y\widetilde{w}^{(\nu)y}_{xx'}\pi_{\mathrm{ss}}(y|x')$.
The largest eigenvalue $\theta_{\mathrm{max}}(\lambda)$ of this matrix is 
\begin{widetext}
\begin{align}
\theta_{\mathrm{max}}(\lambda)=\dfrac{1}{2}\sum_\nu\left[\overline{w}^{(\nu)}_{10}+\overline{w}^{(\nu)}_{01}\right]\left\{-1+\sqrt{1-\dfrac{4\left[\overline{w}^{(L)}_{01}\overline{w}^{(R)}_{10}(1-e^{-\lambda\Delta\mu})+\overline{w}^{(L)}_{10}\overline{w}^{(R)}_{01}(1-e^{\lambda\Delta\mu})\right]}{\left(\sum_\nu\left[\overline{w}^{(\nu)}_{10}+\overline{w}^{(\nu)}_{01}\right]\right)^2}}\right\}.
\end{align}
Then, the second derivative of $\theta_{\mathrm{max}}$ gives the fluctuation $D_W$:
\begin{align}
D_W&=\dfrac{1}{2}\left.\dfrac{\partial^2}{\partial \lambda^2}\mu(\lambda)\right|_{\lambda=0}\notag\\
&=\dfrac{1}{2}\Gamma_X\left.\dfrac{\partial^2}{\partial \lambda^2}\theta_{\mathrm{max}}(\lambda)\right|_{\lambda=0}\notag\\
&=D_n\Delta\mu^2,
% \label{ex2:D_S}
\end{align}
where $D_n$ denotes the fluctuation of the net particle current:
\begin{align}
D_n&=\dfrac{\Gamma_X}{2}\left\{\dfrac{1}{\sum_\nu\left[\overline{w}^{(\nu)}_{10}+\overline{w}^{(\nu)}_{01}\right]}\left[\overline{w}^{(L)}_{01}\overline{w}^{(R)}_{10}+\overline{w}^{(L)}_{10}\overline{w}^{(R)}_{01}\right]-\dfrac{2}{\left(\sum_\nu\left[\overline{w}^{(\nu)}_{10}+\overline{w}^{(\nu)}_{01}\right]\right)^3}\left[\overline{w}^{(L)}_{01}\overline{w}^{(R)}_{10}-\overline{w}^{(L)}_{10}\overline{w}^{(R)}_{01}\right]^2\right\}\notag\\
&=\dfrac{\Gamma_X}{2}\left\{\dfrac{(1-\varepsilon)^2(1-f_L)f_R+\varepsilon^2f_L(1-f_R)}{1+(2\varepsilon-1)(f_L-f_R)}-\dfrac{2\left[(1-\varepsilon)^2(1-f_L)f_R-\varepsilon^2f_L(1-f_R)\right]^2}{\left[1+(2\varepsilon-1)(f_L-f_R)\right]^3}\right\}+O(\tilde{\Gamma}_X).
\end{align}
\end{widetext}

\section{Coupled linear overdamped Langevin equations\label{Appendix: Coupled linear overdamped Langevin equations}}
In this section, we provide a detailed calculation of the fluctuation of the entropy production $D_S$ in the fast relaxation limit of $Y$ for the coupled linear overdamped Langevin equations introduced in Sec.~\ref{Coupled linear overdamped Langevin equations}.
The fluctuation of the information flow $D_I$ can be calculated in a similar way.
We also prove that $D^{\mathrm{II}}_{\mathcal{J}}:=\lim_{\mathcal{T}\rightarrow\infty}\mathrm{Var}[\hat{\mathcal{J}}^{\mathrm{II}}_{\mathcal{T}}]/2\mathcal{T}\rightarrow0$ in the fast relaxation limit of $Y$ for the generalized current $\hat{\mathcal{J}}_{\mathcal{T}}$ whose weight function satisfies the condition (\ref{ex2: condition}).

\subsection{Calculation of $D_S$}
The fluctuation of the stochastic medium entropy production is defined as
\begin{align}
D_S:=\lim_{\mathcal{T}\rightarrow\infty}\dfrac{1}{2\mathcal{T}}\mathrm{Var}[\Delta\hat{S}^X_{\mathrm{env}}],
\end{align}
where $\Delta\hat{S}^X_{\mathrm{env}}$ denotes the stochastic medium entropy production:
\begin{align}
\Delta\hat{S}^X_{\mathrm{env}}=\int^{\tau=\omega^X\mathcal{T}}_{\tau=0}g(x_\tau,y_\tau)\circ dx_\tau,
% &=\int^{t=\mathcal{T}}_{t=0}\dfrac{1}{D^X}(-\omega^X x_t+\omega^{XY} y_t)\circ dx_t\notag\\
\end{align}
where we have used the dimensionless slow time $\tau=\omega^Xt$, and the weight function is defined as
\begin{align}
g(x,y):=\dfrac{1}{\bar{D}^X}(-x+\bar{\omega}^{XY} y).
\end{align}
The fluctuation $D_S$ can be obtained from the scaled cumulant generating function defined by
\begin{align}
\mu(\lambda)=\lim_{\mathcal{T}\rightarrow\infty}\dfrac{1}{\mathcal{T}}\ln\langle e^{\lambda\Delta\hat{S}^X_{\mathrm{env}}}\rangle.
\end{align}
To compute the scaled cumulant generating function, we introduce the generating function conditioned to an initial state $(x_0,y_0)=(x,y)$, defined by
\begin{align}
G_{\mathcal{T}}(x,y):=\langle e^{\lambda\Delta\hat{S}^X_{\mathrm{env}}}|x,y\rangle.
\end{align}
The time evolution of $G_{\mathcal{T}}$ is described by the Feynman-Kac formula~\cite{touchette2018introduction}:
\begin{align}
\partial_\tau G_{\tau}(x,y)=\mathcal{L}^\dag_\lambda[G_{\tau}](x,y),
\label{Feynman-Kac formula_XY}
\end{align}
where $\mathcal{L}^\dag_\lambda$ denotes the tilted generator defined by
\begin{align}
\mathcal{L}^\dag_\lambda=\mathcal{L}^{X\dag}_\lambda+\dfrac{1}{\epsilon}\mathcal{L}^{Y\dag}_\lambda
\end{align}
with
\begin{align}
\mathcal{L}^{X\dag}_\lambda&:=\bar{F}^X(x,y)\left[\partial_x+\lambda g(x,y)\right]+\bar{D}^X\left[\partial_x+\lambda g(x,y)\right]^2,\\
\mathcal{L}^{Y\dag}_\lambda&:=\bar{F}^Y(x,y)\partial_y+\bar{D}^Y\partial^2_y,
\end{align}
where $\bar{F}^X(x,y):=-x+\bar{\omega}^{XY} y$ and $\bar{F}^Y(x,y):=\bar{\omega}^{YX} x-y$ denote the dimensionless drift terms.
The largest eigenvalue of this tilted generator gives the scaled cumulant generating function.

Since we are interested in the fast relaxation limit of $Y$, we can further simplify the problem by considering the effective tilted generator for $X$, as follows.
We first assume that $G_\tau$ have asymptotic expansions in terms of the asymptotic sequences $\{\epsilon^n\}^\infty_{n=0}$ as $\epsilon\rightarrow0$:
\begin{align}
G_\tau=G^{(0)}_\tau+\epsilon G^{(1)}_\tau+\cdots.
\end{align}
Here, we impose the normalization condition
\begin{align}
\int dy\pi_{\mathrm{ss}}(y|x)G^{(0)}_\tau(x,y)&=\int dy\pi_{\mathrm{ss}}(y|x)G_\tau(x,y)\notag\\
&=:G^X_\tau(x),
\end{align}
where $\pi_{\mathrm{ss}}$ denotes the zero-eigenfunction for $\mathcal{L}^Y_0$.
By substituting this expansion into (\ref{Feynman-Kac formula_XY}), we find that the leading order gives
\begin{align}
\mathcal{L}^{Y\dag}_\lambda[G^{(0)}_\tau](x,y)=0.
\end{align}
Since $\mathcal{L}^{Y\dag}_\lambda=\mathcal{L}^{Y\dag}_0$, the zero-eigenfunction for $\mathcal{L}^{Y\dag}_\lambda$ is $1$.
From the Perron-Frobenius theorem and the normalization condition, we find that $G^{(0)}_\tau$ has the form
\begin{align}
G^{(0)}_\tau(x,y)=G^X_\tau(x).
\end{align}
The subleading order of (\ref{Feynman-Kac formula_XY}) gives
\begin{align}
\partial_\tau G^{(0)}_{\tau}(x,y)=\mathcal{L}^{X\dag}_\lambda[G^{(0)}_\tau](x,y)+\mathcal{L}^{Y\dag}_\lambda[G^{(1)}_\tau](x,y).
\end{align}
From the solvability condition for $G^{(1)}_\tau(x,y)$, we obtain the effective dynamics for $G^X_\tau(x)$:
\begin{align}
\partial_\tau G^X_{\tau}(x)=\overline{\mathcal{L}}^{X\dag}_\lambda[G^X_\tau](x).
\end{align}
Here, $\overline{\mathcal{L}}^{X\dag}_\lambda$ denotes the effective tilted generator defined by
\begin{widetext}
\begin{align}
\overline{\mathcal{L}}^{X\dag}_\lambda&:=\int dy\pi_{\mathrm{ss}}(y|x)\mathcal{L}^{X\dag}_\lambda\notag\\
% &=\int dy\pi_{\mathrm{ss}}(y|x)\left\{\bar{F}^X_\tau(x,y)\left[\partial_x+\lambda g(x,y)\right]+\bar{D}^X\left[\partial_x+\lambda g(x,y)\right]^2\right\}.
&=-(1-\bar{\omega}^{XY}\bar{\omega}^{YX})x\partial_x+\bar{D}^X\partial^2_x\notag\\
&\quad+\lambda\left[\dfrac{\bar{D}^Y}{\bar{D}^X}(\bar{\omega}^{XY})^2+\dfrac{1}{\bar{D}^X}(1-\bar{\omega}^{XY}\bar{\omega}^{YX})^2x^2-1-2(1-\bar{\omega}^{XY}\bar{\omega}^{YX})x\partial_x\right]\notag\\
&\quad+\lambda^2\left[\dfrac{\bar{D}^Y}{\bar{D}^X}(\bar{\omega}^{XY})^2+\dfrac{1}{\bar{D}^X}(1-\bar{\omega}^{XY}\bar{\omega}^{YX})^2x^2\right].
\end{align}
We first note that when $\lambda=0$, the largest eigenvalue is $0$ with the corresponding eigenfunction $\phi_{\lambda=0}(x)=1$.
To find the largest eigenvalue for general $\lambda$, we impose the Gaussian ansatz $\phi_\lambda(x)=\exp(- K(\lambda)x^2/2)$.
Then, the largest eigenvalue $\theta_{\mathrm{max}}(\lambda)$ should satisfy
\begin{align}
\theta_{\mathrm{max}}(\lambda)&=\dfrac{\mathcal{L}^\dag_\lambda\phi_\lambda(x)}{\phi_\lambda(x)}\notag\\
&=(1-\bar{\omega}^{XY}\bar{\omega}^{YX})x^2 K(\lambda)+\bar{D}^X(- K(\lambda)+K^2(\lambda)x^2)\notag\\
&\quad+\lambda\left[\dfrac{\bar{D}^Y}{\bar{D}^X}(\bar{\omega}^{XY})^2+\dfrac{1}{\bar{D}^X}(1-\bar{\omega}^{XY}\bar{\omega}^{YX})^2x^2-1+2(1-\bar{\omega}^{XY}\bar{\omega}^{YX})x^2 K(\lambda)\right]\notag\\
&\quad+\lambda^2\left[\dfrac{\bar{D}^Y}{\bar{D}^X}(\bar{\omega}^{XY})^2+\dfrac{1}{\bar{D}^X}(1-\bar{\omega}^{XY}\bar{\omega}^{YX})^2x^2\right].
\end{align}
Because this relation holds for arbitrary $x$, comparing the coefficients of the quadratic form yields
\begin{align}
\theta_{\mathrm{max}}(\lambda)=-\bar{D}^X K(\lambda)+\lambda\left[\dfrac{\bar{D}^Y}{\bar{D}^X}(\bar{\omega}^{XY})^2-1\right]+\lambda^2\dfrac{\bar{D}^Y}{\bar{D}^X}(\bar{\omega}^{XY})^2
\end{align}
and
\begin{multline}
(1-\bar{\omega}^{XY}\bar{\omega}^{YX}) K(\lambda)+\bar{D}^XK^2(\lambda)+\lambda\left[\dfrac{1}{\bar{D}^X}(1-\bar{\omega}^{XY}\bar{\omega}^{YX})^2+2(1-\bar{\omega}^{XY}\bar{\omega}^{YX}) K(\lambda)\right]+\lambda^2\dfrac{1}{\bar{D}^X}(1-\bar{\omega}^{XY}\bar{\omega}^{YX})^2=0.
\label{condition for G}
\end{multline}
\end{widetext}
We now expand $K$ in terms of $\lambda$ as
\begin{align}
 K(\lambda)=\lambda K^{(1)}+\lambda^2K^{(2)}+\cdots.
\end{align}
Here, note that $ K(\lambda)$ should go to zero as $\lambda\rightarrow0$, because $\theta_{\mathrm{max}}(\lambda)$ is the largest eigenvalue and $\theta_{\mathrm{max}}(\lambda)\rightarrow0$ as $\lambda\rightarrow0$.
Then, by substituting this expansion into (\ref{condition for G}), we find that the leading order yields
\begin{align}
(1-\bar{\omega}^{XY}\bar{\omega}^{YX})K^{(1)}+\dfrac{1}{\bar{D}^X}(1-\bar{\omega}^{XY}\bar{\omega}^{YX})^2=0.
\end{align}
From this equation, we obtain
\begin{align}
K^{(1)}=-\dfrac{1}{\bar{D}^X}(1-\bar{\omega}^{XY}\bar{\omega}^{YX}).
\end{align}
The subleading order of (\ref{condition for G}) gives
\begin{multline}
(1-\bar{\omega}^{XY}\bar{\omega}^{YX})K^{(2)}+\bar{D}^X(K^{(1)})^2+2(1-\bar{\omega}^{XY}\bar{\omega}^{YX})K^{(1)}\\
+\dfrac{1}{\bar{D}^X}(1-\bar{\omega}^{XY}\bar{\omega}^{YX})^2=0.
\end{multline}
Since $1-\bar{\omega}^{XY}\bar{\omega}^{YX}>0$, we find that $K^{(2)}=0$.
Therefore, the largest eigenvalue is
\begin{align}
\theta_{\mathrm{max}}(\lambda)=\lambda\left[\dfrac{\bar{D}^Y}{\bar{D}^X}(\bar{\omega}^{XY})^2-\bar{\omega}^{XY}\bar{\omega}^{YX}\right]+\lambda^2\dfrac{\bar{D}^Y}{\bar{D}^X}(\bar{\omega}^{XY})^2.
\end{align}
From this result, we can calculate $D_S$ as
\begin{align}
D_S&=\dfrac{1}{2}\left.\dfrac{\partial^2}{\partial \lambda^2}\mu(\lambda)\right|_{\lambda=0}\notag\\
&=\dfrac{1}{2}\omega^X\left.\dfrac{\partial^2}{\partial \lambda^2}\theta_{\mathrm{max}}(\lambda)\right|_{\lambda=0}\notag\\
&=\omega^X\dfrac{\bar{D}^Y}{\bar{D}^X}(\bar{\omega}^{XY})^2.
% \label{ex2:D_S}
\end{align}

\subsection{Proof of $D^{\mathrm{II}}_{\mathcal{J}}\rightarrow0$\label{Proof for linear weight function}}
Here, we prove that $D^{\mathrm{II}}_{\mathcal{J}}:=\lim_{\mathcal{T}\rightarrow\infty}\mathrm{Var}[\hat{\mathcal{J}}^{\mathrm{II}}_{\mathcal{T}}]/2\mathcal{T}=0$ in the fast relaxation limit of $Y$ for the generalized current $\hat{\mathcal{J}}_{\mathcal{T}}$ whose weight function is given by $g(x,y)=C(y-\bar{\omega}^{YX}x)$.
For this weight function, $\hat{\mathcal{J}}^{\mathrm{II}}_{\mathcal{T}}$ can be expressed as
\begin{align}
\hat{\mathcal{J}}^{\mathrm{II}}_{\mathcal{T}}&:=\int^{\omega^X\mathcal{T}}_0f(x_\tau,y_\tau)d\tau,
\end{align}
where
\begin{multline}
f(x,y):=C(y-\bar{\omega}^{YX}x)(-x+\bar{\omega}^{XY} y)\\
+\bar{D}^X\partial_xC(y-\bar{\omega}^{YX}x).
\end{multline}
The fluctuation $D^{\mathrm{II}}_{\mathcal{J}}$ can be obtained from the following scaled cumulant generating function,
\begin{align}
\mu(\lambda)=\lim_{\mathcal{T}\rightarrow\infty}\dfrac{1}{\mathcal{T}}\ln\langle e^{\lambda\hat{\mathcal{J}}^{\mathrm{II}}_{\mathcal{T}}}\rangle,
\end{align}
which corresponds to the largest eigenvalue of the following tilted generator~\cite{touchette2018introduction}:
\begin{align}
\mathcal{L}^\dag_\lambda=\mathcal{L}^{X\dag}_\lambda+\dfrac{1}{\epsilon}\mathcal{L}^{Y\dag}_\lambda
\end{align}
with
\begin{align}
\mathcal{L}^{X\dag}_\lambda&:=\bar{F}^X(x,y)\partial_x+\bar{D}^X\partial^2_x+\lambda f(x,y),\\
\mathcal{L}^{Y\dag}_\lambda&:=\bar{F}^Y(x,y)\partial_y+\bar{D}^Y\partial^2_y.
\end{align}
The effective tilted generator is then given by
\begin{align}
\overline{\mathcal{L}}^{X\dag}_\lambda&:=\int dy\pi_{\mathrm{ss}}(y|x)\mathcal{L}^{X\dag}_\lambda\notag\\
% &=\int dy\pi_{\mathrm{ss}}(y|x)\left\{\bar{F}^X_\tau(x,y)\left[\partial_x+\lambda g(x,y)\right]+\bar{D}^X\left[\partial_x+\lambda g(x,y)\right]^2\right\}.
&=-(1-\bar{\omega}^{XY}\bar{\omega}^{YX})x\partial_x+\bar{D}^X\partial^2_x\notag\\
&\qquad+\lambda C\left(-\bar{\omega}^{YX}\bar{D}^X+\bar{\omega}^{XY}\bar{D}^Y\right).
\end{align}
By performing a similar calculation as in the previous section, we finally obtain 
\begin{align}
\theta_{\mathrm{max}}(\lambda)=\lambda C\left(-\bar{\omega}^{YX}\bar{D}^X+\bar{\omega}^{XY}\bar{D}^Y\right).
\end{align}
Thus, we find that $D^{\mathrm{II}}_{\mathcal{J}}:=\lim_{\mathcal{T}\rightarrow\infty}\mathrm{Var}[\hat{\mathcal{J}}^{\mathrm{II}}_{\mathcal{T}}]/2\mathcal{T}=0$.

\bibliography{bipartite_TUR}

\end{document}